\documentclass[12pt]{article}
\usepackage[margin = 1in]{geometry}
\usepackage{setspace}
\usepackage{amsfonts}
\usepackage{amsmath}
\usepackage{graphicx}
\usepackage{xcolor}
\usepackage[
backend=biber,
style= apa,
sorting=nyt, 
maxcitenames = 1,
mincitenames=1,
uniquename = false, 
natbib=true
]{biblatex}
\usepackage[hidelinks,colorlinks=true,linkcolor=blue,citecolor=blue]{hyperref}
\usepackage{soul}
\usepackage{hyperref}
\usepackage{booktabs}
\usepackage{subcaption}

\begin{document}

\title{Outcome Modeling in Design-Based Inference for Spatial Settings}

\author{Arisa Sadeghpour \thanks{Department of Statistics, UC Berkeley. This research was supported in part by AFOSR MURI grant \#FA9550-22-1-0380.}\and Erin Hartman \thanks{Department of Political Science, UC Berkeley.  This research was supported in part by NSF grant SCC-2125319.\\ The authors would like to thank Jeff Brantingham, Peng Ding, Avi Feller, Melody Huang, Sam Pimentel, Ye Wang, and members of the Berkeley Causal Lab. They would also like to thank Santiago Tob\'on for kindly sharing data.}}

\date{\today}

\begin{titlepage}
\maketitle

\vspace{0.5cm}
\onehalfspacing
In the face of spatial interference, researchers are often interested in estimating treatment effects at specific points located in space. \citet{Wang2025aoas} and \citet{pollmann2023causalinferencespatialtreatments} provide design-based frameworks for estimating spillover effects on points located across a range of distances from interventions. Although these frameworks are design-based, we show their proposed estimands rely on outcomes that are directly unobservable and therefore, require outcome modeling. When using modeled outcomes in practice, even the typically design-unbiased Horvitz-Thompson estimator can accrue bias as a result of the modeling error. The performance of spatial outcome models depends on the density or resolution of observed outcomes. Through simulation, we find that the bias of the estimators decays with increasing outcome density, but not with increasing numbers of intervention units, and standard errors using modeled outcomes converge to the oracle standard errors. To demonstrate the role of outcome modeling with spatial interference, we reanalyze an experiment from \citet{collazos2021hotspot} on the effect of hot spots policing on crime and provide several suggestions for practice.

\end{titlepage}

\doublespacing
\section{Introduction}

An important question for researchers is often how the impact of an intervention transmits through space. Intervention nodes located in a geographic space are randomly or quasi-randomly assigned to treatment, but we are interested in outcomes and effects measured at locations distinct, and distant, from intervention nodes.  In this setting, treatment effects are often assumed to ``spill over" from intervention nodes throughout the geographic space, rendering the standard stable unit treatment value assumption (SUTVA) untenable because the potential outcome at a specific location depends on the interference of treatment assignments of many intervention nodes.

For example, \citet{collazos2021hotspot} study the effect of hot spots policing on crime in Medellín, Colombia. In addition to direct effects, the authors estimate spillover effects for two types of streets: those within short-range (125m) of treated hot spots, and those within long-range (125m to 250m) of treated hot spots. However, policing can have hyperlocal effects \citep{braga2014hotspots} and criminologists are often interested in how those effects vary with distance from treated hot spots as well as whether there are displacement effects \citep{braga2014hotspots, braga2019hot, weisburd2006does}.  These types of questions are difficult to answer with the coarsened and aggregated analyses typically conducted.  The methods we discuss in this paper are better suited to answering this style of question: what is the impact of hot spots policing over a range of, or at a specific, distances?

We build on two design-based frameworks proposed for the study of spatial experiments of this type that account for interference. \citet{Wang2025aoas} extend the average direct effect of \citet{hudgens2008interference} and the expected average treatment effect from \citet{savje2021interference} to define an “average marginalized effect” (AME). The AME is indexed by distance from an intervention node to estimate the effect of treatment at a given distance, marginalizing over treatment assignments of all intervention nodes. \citet{pollmann2023causalinferencespatialtreatments} studies a similar estimand, but works in a quasi-experimental setting.  Both of these design-based frameworks seek to estimate spillover effects at specific distances away from intervention nodes, as opposed to other common approaches that estimate cumulative or coarsened distance effects.  Moreover, they can both be categorized as interference in \textit{bipartite} settings because intervention nodes are distinct from where outcomes are measured \citep{zigler2021bipartite}. 

What is underexplored in the existing literature is the role of outcome measurement when estimating these effects.  Unlike most causal inference settings, in this scenario, where outcomes are measured at locations distinct from the intervention, and in which we care about the effect at specific distances, researchers must typically estimate outcomes rather than observe them directly.  For example, when using rasterized data, we do not know the exact number of crimes that occur on a 100m radius from an intervention since we typically do not have measurements exactly 100m around every intervention node.  Given these spatial data constraints, we must first estimate the number of crimes around both the treated and control nodes in order to further estimate the effect at 100m.  Our paper highlights that despite existing approaches being design-based, naturally suggesting common design-unbiased estimators, the spatial setting requires additional estimators for the observed potential outcome.  Our paper uncovers this nuance, that spatial causal inference often necessitates outcomes models, and outlines the implications for practice.  

In the following paper, we dive further into this need to employ outcome modeling in spatial interference settings, making three contributions.  First, we show that common design and measurement decisions, paired with outcome modeling decisions, can have a sizable impact on the analysis of a spatial experiment.  We reveal how outcome modeling can impart bias; even the oracle Horvitz-Thompson estimator would be unbiased.  We also highlight the importance of thinking about the number and density of outcome measurements in addition to the number of intervention nodes, i.e. the sample size of the study.   Second, we discuss complications with common model selection approaches, articulating a reasoned approach.  Finally, we demonstrate the impact of these decisions through a novel re-analysis of \citet{collazos2021hotspot} by estimating the impact of hot spots policing as a function of distance.

\subsection{Application to hot spots policing in Medellín}

We will re-analyze \citet{collazos2021hotspot}, who study the impact of hot spots policing in Medellín.  We describe the original study design here.  Out of 37,055 street segments in Medellín, \citet{collazos2021hotspot} identified 967 hot spots, defined as the approximately 3\% of streets with the highest pre-treatment (baseline) crime outcomes, which we refer to as intervention nodes. 384 of these intervention nodes, or hot spots, were randomly assigned to a six-month increase in police patrols from May to November of 2015. Prior to the intervention, each hot spot received roughly one hour of patrolling a day. The intervention increased patrolling to at least 105 minutes per day divided into 7 entries of 15 minutes each. \citet{collazos2021hotspot} did not employ simple random assignment, but rather randomly assigned streets to treatment or control following a set of restrictions imposed by the Metropolitan Police. Therefore, the probability of treatment was not equal across streets, which we account for in our reanalysis.  Our primary observed outcome is a street-level\footnote{We assume street segments are short enough such that they can be considered point interventions. Alternatively, if we had access to the shapes of the street segments, we could consider them as polygon interventions and use the associated estimation strategy proposed by \citet{Wang2025aoas}.}, aggregated crime index as defined in \citet{collazos2021hotspot} and later used in \citet{puelz2022randomization}.

\begin{figure}[ht]
    \centering
    \includegraphics[width=0.49\linewidth]{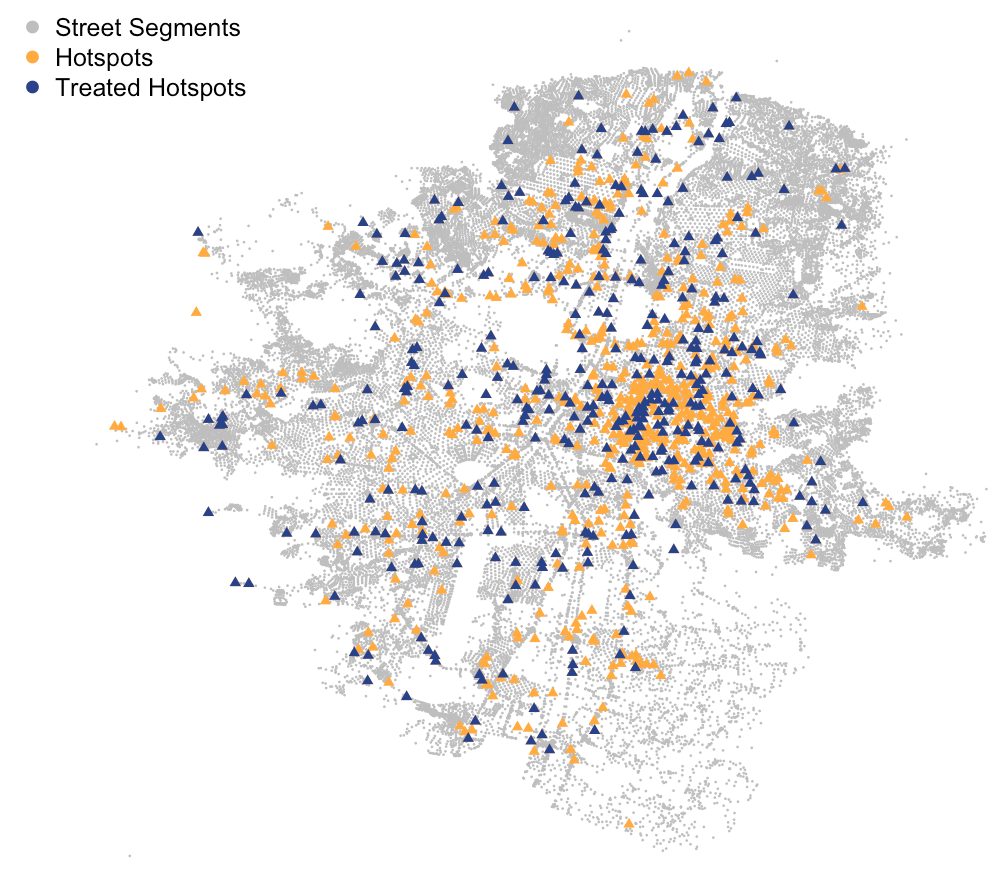}
    \caption{Crime measurements at street segments in Medell\'{i}n with intervention nodes overlaid.}
    \label{fig:medellin-streets }
\end{figure}

\section{Setting and Notation}

In the following section, we build on the work and notation introduced in \citet{Wang2025aoas}.  We summarize the notation here, focusing on key quantities and assuming quantities meet any regularity conditions outlined in their original paper. Let there be $N$ intervention nodes, indexed by $i$, that reside in a two dimensional set $\mathcal{X}$ indexed by $x = (x_1, x_2)$ (e.g. latitude and longitude).\footnote{See \citet{Wang2025aoas} for an extension where intervention nodes are polygons, such as counties.}  Following \citet{Wang2025aoas}, we assume a binary treatment $Z_i$ that follows a Bernoulli randomization with probability of assignment $\pi_i$ for each intervention node.\footnote{With large $N$, Bernoulli assignment is a reasonable approximation for completely randomized designs \citep{savje2021interference}.}  The treatment vector is captured by $\mathbf{Z} \equiv (Z_1, \dots, Z_N)$, the possible values of which are determined by the experimental design, and the realized assignment from the experiment is denoted $\mathbf{z} = (z_1, \dots, z_N) \in \{0, 1\}^N$.  In our application in Medell\'{i}n, intervention nodes are streets that are identified to be hot spots, meaning that they have high levels of baseline crime, some of which are randomly assigned to the hot spots policing condition.  
While we focus on the experimental setting, see \citet{pollmann2023causalinferencespatialtreatments} for extensions of this framework to observational settings, where an exchangeable group of control locations must be identified under an ignorability assumption. 

\subsection{Outcome Measures}
Unlike in unit-level experimental designs where we might only estimate the effect on the intervention nodes themselves, in the spatial setting we can consider the effect at every location in space, $x \in \mathcal{X}$, not just on the set of intervention nodes.  Additionally, due to interference, potential outcomes $Y_x(\mathbf{z})$, which capture the outcome at point $x$ given a realized treatment vector $\mathbf{z}$, are defined for each of the $2^N$ possible  treatment assignment vectors. To summarize the effect at a specific distance, $d$, from intervention nodes, we follow \citet{Wang2025aoas} and define the “circle average”, or the effect averaged over a ``circle" of distance $x$ around nodes.  We define the function:

\begin{equation}
    \mu_i(\mathbf{Y}(\mathbf{z}), \Omega_d) = \frac{\int_{x: d_i(x) \in \Omega_d} Y_x(\mathbf{z}) d\zeta}{\int_{x: d_i(x) \in \Omega_d} d\zeta}
    \label{eq:mu}
\end{equation}

In this expression, $d_i(x)$ measures the distance between point $x$ and intervention node $i$, $\Omega_d$ is a set of distance values, and $\zeta$ is a suitable measure on the space, and thus $\mu_i(\mathbf{Y}(\mathbf{z}), \Omega_d)$ is the average across points with defined distance $d$. The choice of $\Omega_d$ determines the shape over which we average outcomes, and \citet{Wang2025aoas} suggest three shapes. If $\Omega_d$ is a singleton ($\Omega_d = \{d\}$), the shape would be a \textit{circle} centered at the intervention node with radius $d$. The circle measures the spillover effect right at distance $d$, for varying distances. $\Omega_d$ can also be defined to be a \textit{donut} $\{d_i(x) : d-\kappa < d_i(x) \leq d\}$, similar to \citet{pollmann2023causalinferencespatialtreatments}, where a user-chosen $\kappa$ dictates the thickness. Lastly,  $\Omega_d$ can be a \textit{disk}, $\{d_i(x) : d_i(x) \leq d\}$, which is similar to the cumulative effects studied in many spatial experiments (e.g. \citet{collazos2021hotspot}).

\subsection{Estimands and Estimators}
With the definition of the observed circle average, we can define causal quantities of interest.  First, define the unidentifiable individualistic marginalized effect at distance $d$ for intervention node $i$ as $\tau_i(d)  = \mu_i(1,d) - \mu_i(0,d)$, which is the difference in circle averages a distance $d$ away from node $i$ when node $i$ is switched from treatment to control, marginalized over all possible treatment assignments to nodes $-i$. From this, the average marginalized effect (AME) for distance $d$ is the average of the individualistic marginalized effect over the $N$ intervention nodes: 
\begin{equation}
    AME(d) = \frac{1}{N}\sum_{i=1}^N \tau_i(d)
    \label{eq:ame}
\end{equation}
\noindent The AME at $d$ is interpreted as the average treatment effect on points a distance $d$ from intervention nodes, marginalized over possible realizations of treatment assignment.

\citet{Wang2025aoas} outline several assumptions necessary for identification of the AME. In addition to bounded potential outcomes, they assume local interference, meaning that nodes beyond some distance from each other are assumed to not interfere with one another, limiting the extent of interference (see Appendix \ref{sec:local-interference}). Additionally, as $N$ grows, the number of nodes that are in proximity of each other is uniformly bounded. This intervention node spacing assumption is met, for example, when the geographic space is first divided into disjoint areas, and intervention nodes are placed in each area. Under these assumptions, the AME is identified under random assignment of treatment to the intervention nodes.

Additionally, the Horvitz-Thompson estimator of the circle averages (Equation ~\eqref{ht-eqn}), when $\mu_i(\mathbf{Y}; d)$ is known, is design-unbiased and consistent in the number of intervention nodes, and the H\'ajek estimator is consistent.

\begin{equation}
\label{ht-eqn}
    \hat\tau_{HT}(d) = \frac{1}{N}\sum_{i=1}^N \frac{Z_i \mu_i(\mathbf Y, d)}{\pi_i} -\frac{1}{N}\sum_{i=1}^N \frac{(1-Z_i) \mu_i(\mathbf Y, d)}{1-\pi_i}
\end{equation}

\section{The Necessity of Outcome Modeling}

In most applied settings, the circle averages, $\mu_i$, are not directly observable. This fact is, of course, known to applied researchers and addressed, for example, in the application of \citet{Wang2025aoas}, but it is not clearly acknowledged in the development of the original results. To demonstrate this point, we focus on the most common setting where $\Omega_d$ is a circle and geostatistical outcomes, $Y_x$, represent an attribute of location $x$ (see Section \ref{sec:alt_outcomes} for discussion on other shapes and outcome types). For example, in the crime setting, $Y_x$ represents the underlying, continuous crime risk at each location as often conceptualized by criminologists \citep{brantingham201721, caplan2015risk}. As seen from Equation ~\eqref{eq:mu}, circle averages are constructed by averaging outcomes $Y_x$ at all (infinitely many) locations along the circle. In practice with finite data, however, $Y_x(\mathbf{z})$ can only be observed for a discrete set of locations, $x \in \mathcal{X}_o \subset \mathcal{X}$. This means that the circle averages are unobserved and must be estimated in order to estimate the AME. 
 
\begin{figure}[ht]
    \centering
    \begin{subfigure}[b]{0.34\linewidth}
        \includegraphics[width=\linewidth]{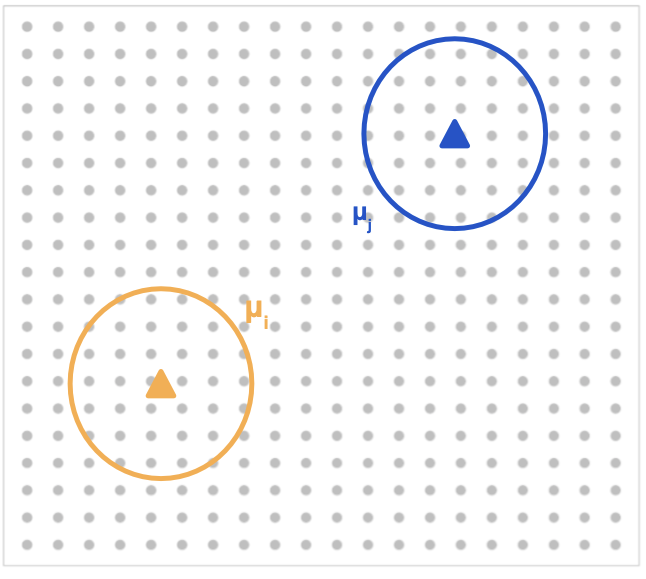}
        \caption{Outcomes arranged on grid. }
        \label{fig:grid-circles}
    \end{subfigure}
    \hspace{0.05\linewidth}
    \begin{subfigure}[b]{0.28\linewidth}
        \includegraphics[width=\linewidth]{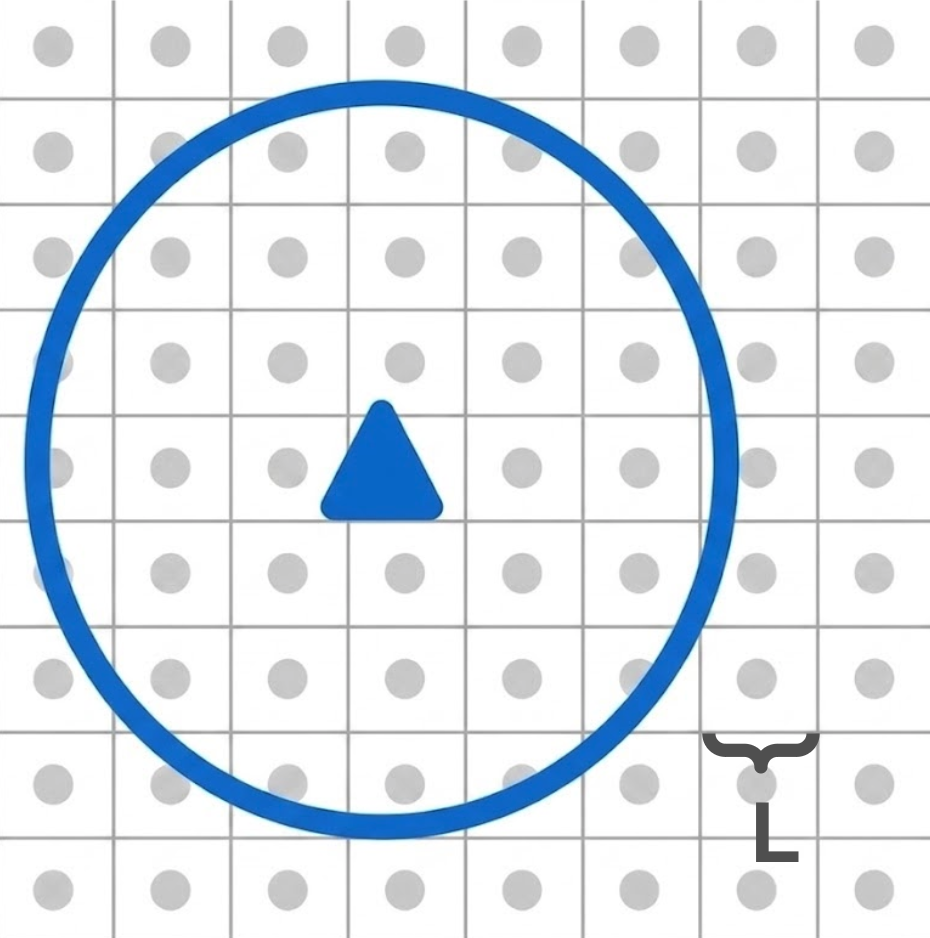}
        \caption{Discretized model with grid size, $L$. }
        \label{fig:discretized}
    \end{subfigure}
    \caption{Two nodes are represented by triangles, and observed outcomes are represented by gray dots. Circle averages $\mu_i$ and $\mu_j$ are shown for one radius, $R$.}
    \label{fig:grid}
\end{figure}

Estimating circle averages can be complicated due to measurement constraints with spatial data. If researchers could observe outcomes at a random sample of points along the desired circle, the circle averages would be unbiased. However, spatial data is often arranged on a grid or raster as in Figure \ref{fig:grid-circles}, so observed outcomes may not lie on or even near circles. Observed outcomes must be first used to construct an outcome model. Then, estimated outcomes for random sample of points along the circle can be used construct the estimated circle averages, $\hat\mu_i$ \citep{samii2023spatialeffect}. This procedure often results in estimation error in the circle averages, and these errors propagate to the AME estimator. 

When outcomes are observed on a grid, it is natural to exploit the data structure and model outcomes using what we refer to as the \textit{discretized model} or \textit{nearest neighbor interpolant}. Each observed point lies in the center of a square raster cell and the outcome of a desired location $x$ is assumed to be the value of the raster cell in which $x$ is located (see Figure \ref{fig:discretized}). Since the discretized model is a simple interpolator common with spatial data, we analytically characterize the errors in the circle averages under this model in Section \ref{sec:discretized_example}.
Section \ref{sec:alt_models} discusses other possible models.

 Several factors impact circle average estimation. The first is the \textit{outcome density} or \textit{resolution}, which refers to the number of observed outcomes per unit area. Intuitively, as density increases, we expect outcome models to perform better and errors in the circle averages to decrease. In the spatial statistics literature, this idea of increasing outcome density is known as \textit{infill} \citep{cressie2015statistics} or \textit{fixed-domain} \citep{stein1999interpolation} asymptotics. Under the discretized model, density is inversely related to the length of the raster cell, $L$. That is, with more observed outcomes per unit area, each square cell decreases in length.  Section \ref{sec:discretized_example} formalizes infill convergence rates with the discretized model.

When designing spatial experiments, another consideration is how the $N$ nodes should be placed relative to the grid. Nodes may be \textit{systematically} placed or (plausibly exogenously) \textit{randomly} placed. We refer to systematically located nodes as those arranged in a grid with some fixed spacing, such as might occur when nodes align with a city's street grid. On the other hand, randomly located nodes have no pattern or constraints to where they are located, up to the constraints given by \citet{Wang2025aoas} on intervention node spacing. In the next section, we formally demonstrate the impact of node placements on circle average estimation under the discretized model. 

Ultimately, the estimation error in the circle averages propagates to error in the AME. Even the typically design-unbiased Horvitz-Thompson estimator accrues bias as a result of the modeling error.  Without imposing any assumptions on the error, we can express the estimated circle average $\hat\mu_i(Y, d)$ as $\mu_i(Y, d)  + \tilde\epsilon_i(Y, d) $. That is, $\tilde\epsilon_i$ denotes the error in the estimated circle averages (as opposed to the error in a particular point).  Then, the bias in the Horvitz-Thompson estimator is: 
\begin{equation}\label{eq:HT_bias}
    \text{bias}(\hat{\tau}_{HT}(d)) = \mathbb{E}_Z\left\{ \frac{1}{N}\sum_{i=1}^{N} \frac{Z_i \, \tilde{\epsilon}_i(Y(z),d)}{\pi_i} - \frac{1}{N}\sum_{i=1}^{N} \frac{(1-Z_i)\, \tilde{\epsilon}_i(Y(z),d)}{1-\pi_i} \right\}
\end{equation}
As is clear in the equation above, the bias in the AME Horvitz-Thompson estimator inherits the same properties as the modeling error of the circle averages, which is our focus.

\section{Understanding Modeling Errors: The Discretized Model}\label{sec:discretized_example}

To provide more intuition about how modeling errors impact causal effect estimation, we provide an analytical investigation of the discretized model described above.  We summarize the results here; more details can be found in Appendix~\ref{app:theory}.  Rasterized data, in which each observed point lies in the center of a square raster cell of length $L$, and the outcome of a location $x$ is assumed to be the value of the raster cell in which $x$ is located (see Figure \ref{fig:discretized}), is common in spatial data.  We investigate the performance of alternative models in our simulations in Section~\ref{sec:sim_setup}.  In our simulated data-generating process, the results provided in this section are similar for the alternative outcome models.

 Although the AME is constructed using several circle averages, to illustrate the main theoretical argument, we focus on estimation of one circle average at distance $d$ under an arbitrary realized treatment assignment.\footnote{For simplicity, we drop the realized treatment assignment $\mathbf{z}$ from the notation in this section.} Consider the circle of radius $d$ centered at intervention node $c \in \mathbb{R}^2$. For any point $x \in \mathbb{R}^2$, denote the distance to the node $c$ as $d_x = ||x - c||_2$.  In a design-based framework, the outcome $Y_x$ at point $x$ is fixed. The outcome is an attribute of the location, and we represent it as the sum $Y_x = f(x) + \epsilon_x$. The effect data generating process, $f(x)$, is a smooth and bounded function. We will assume, without loss of generality, that $f(x)$ only depends on the distance between the point $x$ and the node, $c$. \footnote{That is, for two points $x_i$ and $x_j$ such that $d_{x_i} = d_{x_j}$, $f(x_i) = f(x_j)$. In other words, although $f$ is defined by the Cartesian coordinate system, in fact, only the radial distance to the node matters.} $\epsilon_x$ is a fixed, idiosyncratic error that can be thought of as a realized draw from a mean-zero distribution. We assume errors are mutually uncorrelated across locations. 

First, we consider the true circle average.  Since only radial distance to the node matters, we represent the point $x$ by its polar representation, $ c + r(\theta, d)$ where $r(\theta, d) = \begin{pmatrix} d\cos(\theta) \\ d\sin(\theta) \end{pmatrix}$. Then, for a circle centered at $c$ with radius $d$, the true circle average $\mu(c, d)$ is:
\begin{equation}
    \mu(c, d) =  \frac{1}{2\pi} \int_0^{2\pi}  f(c + r(\theta, d)) \, d\theta
\end{equation}

Now we consider the estimated circle average.  Define a grid in $\mathbb{R}^2$ with each cell being a square of length $L$. Each cell $k$ has a center or midpoint with coordinates $g_k \in \mathbb{R}^2$.  Under the discretized model, the value at any point $p$ is modeled as the value of the grid it falls in. A grid takes the value of the center of the grid, $g_k$. In other words, the outcome at a point $p$ is estimated as the outcome at the $g_k$ nearest it. The discretized model, $D(p)$, is then: 
\begin{equation}
    D(p) := Y(g_k) = f(g_k) + \epsilon_{g_k} \quad \text{for } p \in \text{Cell k}
\end{equation}
A point $p$ on the circle is offset from the nearest grid center by $\delta$:
\begin{equation}
    g_k = p + \delta \quad \text{with } \delta \in \left[-\frac{L}{2}, \frac{L}{2}\right] \times \left[-\frac{L}{2}, \frac{L}{2}\right]
\end{equation}

\begin{figure}[h]
    \centering
    \includegraphics[width=0.5\linewidth]{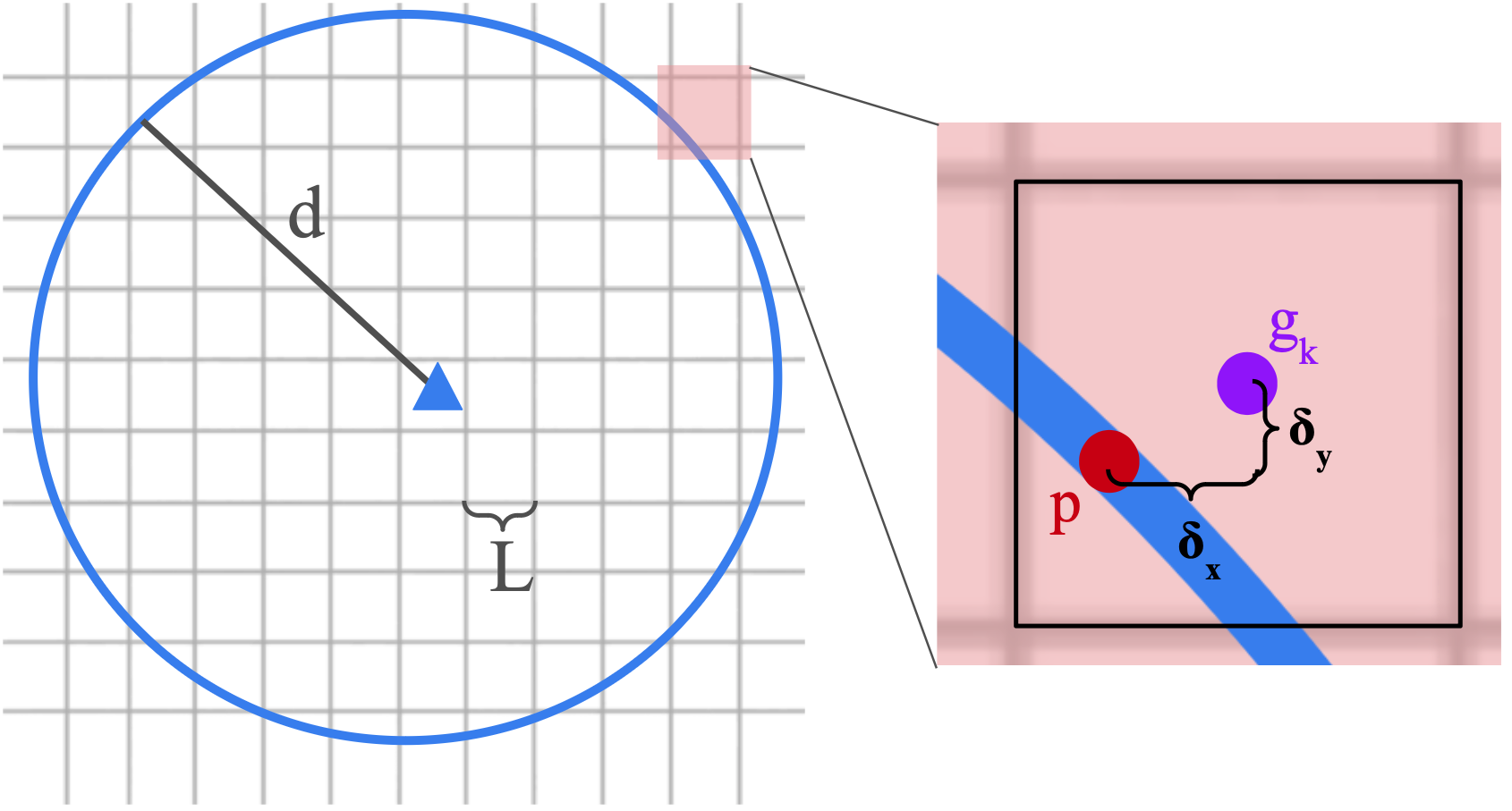}
    \caption{Diagram showing overall setup, point of interest $p$, center of cell $g_k$, and offset. \label{fig:offsets}}
\end{figure}

The magnitude of the offset can be at most $\frac{L}{2}$ in both the $x$ and $y$ direction since the cell is of size $L$. We use $\delta_x$ and $\delta_y$ to denote the $x$ and $y$ components of the offset, shown in Figure \ref{fig:offsets}.  

The circle average estimator using the discretized model is:
\begin{equation}
    \hat{\mu}(c, d) = \frac{1}{2\pi} \int_0^{2\pi} D(c + r(\theta, d)) \, d\theta
\end{equation}

Finally, the estimation error in the circle average is the difference between the estimated and true circle averages:

\begin{equation}
    \tilde{\epsilon}(d) := \hat{\mu}(c, d) - \mu(c, d) = \frac{1}{2\pi} \int_0^{2\pi}  D(c + r(\theta, d)) - f(c + r(\theta, d))\, d\theta 
\end{equation}

For a fixed $L$, this estimation error is not generally zero. In Appendix \ref{app:theory} we use a Taylor expansion argument to characterize the estimation error as a function of $L$ when $d \gg L$. In particular, we show that the estimation error of a circle average is in the worst case $O(L)$ (see Appendix \ref{app:sys_theory}), meaning that the estimation error will disappear under infill asymptotics.  Furthermore with randomly placed nodes, we further show that the expected estimation error across node placements is $O(L^2)$ (see Appendix \ref{app:rand_theory}). That is, by randomly placing nodes, the first order term in the Taylor expansion disappears and the convergence rate in terms of infill asymptotics is faster than with systematically placed nodes. 

\begin{figure}[ht]
    \centering
    \includegraphics[width=0.2\linewidth]{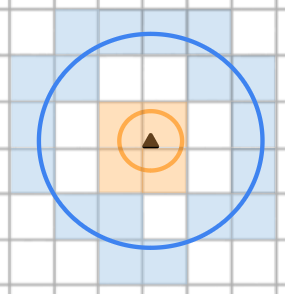}
    \caption{Estimation of circle averages at larger distances uses more points compared to smaller distances. The orange circle with a smaller radius lies in only four cells, while the blue circle lies in 18.}
    \label{fig:vary-d}
\end{figure}

Estimating circle averages at small distances $d$ is challenging because there are fewer outcomes near the circle.  This is particularly problematic for applied researchers, because effects are often largest and of most substantive interest at small distances. As shown in Figure \ref{fig:vary-d}, as the radius increases, each circle passes through more distinct cells and the average distance between each circle and the cell center approaches zero.  This means that bias will decrease as $d$ increases.  Appendix \ref{app:theory_decay} formalizes this intuition.  An implication of this is that, despite systematically placed nodes having more estimation error than randomly placed nodes, with large $d$ this issue is mitigated.

Ultimately, the question is how the estimation error in the circle averages impacts our causal estimates.  Turning back to Equation~\eqref{eq:HT_bias}, we have shown that the estimation error for a single circle average, $\tilde\epsilon_i(\mathbf{Y}(\mathbf{z}),d)$, is generally not zero. The estimation error in the circle averages induces  bias even for the typically design-unbiased Horvitz-Thompson estimator. The specific bias will depend on the data-generating process and node placements, which we have not defined here. Moreover, with estimated circle averages, neither the Horvitz-Thompson nor the H\'ajek estimator are guaranteed to remain consistent in the number of intervention nodes; only under infill asymptotics are they consistent.

The problem we have described is an example of measurement error in the outcome, in this case the outcome being the circle averages. If errors are classical under a measurement model, meaning they are mean-zero and uncorrelated with the true circle average and treatment status (i.e. non-differential), it is well known that the Horvitz-Thompson estimator remains design-unbiased, although variance increases \citep{sarndal2003model}. In most spatial settings, however, errors are not classical. Appendix \ref{app:theory} shows that errors depend on the gradients of effects emanating from nodes. When a true AME effect exists, these gradients differ across treatment conditions, causing errors to be non-classical. The resulting bias can be seen empirically in Section \ref{sec:sim_results}. Some exceptions exist, particularly, when gradients are non-differential across treatment status. For example, when there is no effect, errors are classical because gradients, and thus errors, are uncorrelated with treatment status. Additional simulations in Appendix \ref{otherdgp} show that in this case the Horvitz-Thompson estimator remains unbiased. 

This section shows that, under the common rasterized data framework, we expect a discretized model to have bias in the estimation of the circle averages, and thus in the causal estimand of interest, the AME.  We have shown that the discretized model is consistent under infill asymptotics, although the rates will be better when nodes are randomly placed than when they are systematically placed.  Bias also decreases with larger radii in the circle average.  In Appendix~\ref{app:theory} we provide formal results and discuss the required underlying assumptions on the DGP.

\section{Alternative outcome types and models}\label{sec:alt_outcomes}

\subsection{Other types of outcomes }
We previously motivated the problem with geostatistical outcomes which represent inherent attributes of the location $x$ (e.g. crime risk at location $x$). Alternatively, researchers may use other types of outcomes, $Y_x$ \citep{cressie2015statistics}. Outcomes at location $x$ can represent aggregate (or rasterized) information (e.g. the number of crimes that occurred in a city block where $x$ is located) often referred to as areal or lattice data in the spatial statistics literature.\footnote{We assume the unit of aggregation is fixed. However, we acknowledge the choice of the unit of aggregation can impact statistical analyses, known as the Modifiable Areal Unit Problem \citep{openshaw1979million, fotheringham1991modifiable}.} Lastly, outcomes may represent the occurrence of an event, such as if a crime happened at location $x$. For this outcome type, we assume all events are observed and view outcomes as arising within a spatial point process framework. We will refer to these three possibilities as \textit{attribute outcomes, aggregate outcomes, or occurrence outcomes}, respectively.

We believe highlighting the necessity of outcome modeling is crucial because it is needed for almost all outcome types above when considering the AME defined in Equation~\eqref{eq:ame}. The impact may be more pernicious depending on the type of outcome and shape used in estimation.  When considering a true circle average, we will require outcome modeling for all outcome types (attribute, aggregate, and occurrence) as it is impossible to directly observe the circle averages because researchers never have access to outcomes at all points along the circle.  For occurrence outcomes, the probability of an event occurring exactly on the circle is zero, and point process models can only estimate the intensity of events for points along the circle. 

Some exceptions to the necessity of outcome modeling exist, though. With occurrence outcomes, donut or disk averages do not require modeling as they can be directly observed by counting the number of events within the shape. For attribute and aggregate outcomes, estimation will still be necessary but may be less difficult when using aggregate outcomes with disks or donuts. The union of the areas of aggregation will likely never exactly match the donut or the disk of interest; however, it is possible that the area of interest lines up reasonably with the unit of aggregation. We note that the notion of infill asymptotics only holds for attribute and aggregate outcomes and not for occurrence outcomes.\footnote{ Since all events are observed, there is no notion of increasing the number of measurements with occurrence outcomes; researchers cannot collect more data of this type. With aggregate outcomes, increasing density is equivalent to decreasing the size of the unit of aggregation.}

\subsection{Other outcome models}\label{sec:alt_models}

The spatial statistics literature provides many choices for spatial models to estimate outcomes. We focus on the discretized model in our analysis below, but in our simulations and application we will also consider two other common choices: inverse distance weighting and kriging. Inverse distance weighting (IDW) is a deterministic interpolation method where the outcomes of unknown points are estimated with a weighted average of known outcomes for points located near the unknown point, and is similar to a method used in \citet{pollmann2023causalinferencespatialtreatments}. The average is weighted inversely proportional to the distance between the unknown target point and observed outcome points \citep{cressie2015statistics}. Kriging is a statistical interpolation model that takes into account autocorrelation in the data \citep{cressie1990origins}. While we focus on these common models, there are many other choices including spatial lag models, spatial error models, and nearest neighbor interpolation to name a few \citep{cressie2015statistics, moraga2023spatial}. 

The choice of model depends on which type of outcome the researcher observes. Regardless of which model is chosen, circle averages can be estimated by averaging the predicted outcome for many points along the desired circle. Convergence rates under infill asymptotics may differ under each model and are beyond the scope of this paper, although similarly considered in existing literature (see \citet{wang2020prediction} for kriging and \citet{farwig1986rate} for IDW).

\section{Simulation Setup}\label{sec:sim_setup}

While the analysis above shows, analytically, that outcome modeling can result in bias in estimation of causal effects, it is limited to the discretized outcome model.  To understand and compare the performance of multiple common spatial outcome models, we conduct a simulation study, structuring our data-generating process on a setting similar to our analysis of hot spots policing. To that end, we consider a rasterized grid of observed outcomes. Our simulations contain several specifications of other design features researchers may consider, including node placement, number of intervention nodes, and outcome density. The first specification is how nodes are placed which can be systematically or randomly (up to the constraints outlined in \citet{Wang2025aoas}). Note that the AME defined in Equation~\eqref{eq:ame} conditions on the node placement. In simulations, we find the specific placement of nodes relative to the outcomes can greatly impact the estimation of the circle averages (see Appendix \ref{placements}). To account for this variation when nodes are randomly placed nodes, we present an average bias over 50 node placements.  We use a Bernoulli randomization with $\pi_i = 0.5$ for assignment of intervention nodes under both random and systematic placements.

The second specification is the number of intervention nodes, $N$. The performance of the aforementioned design-based Horvitz-Thompson and H\'ajek estimators are consistent with oracle outcome measures in the number of nodes. We simulate experiments with $N \in \{9, 100, 196\}$ to assess this asymptotic behavior when outcomes are estimated. Due to the assumption on intervention node spacing needed for inference of the AME, as the number of nodes increases, so does the geographic space the experiment occupies in order to hold density fixed. 

The third parameter is the density or resolution of observed outcomes. In our simulations, we explore the impact of having low, medium, and high density of observed outcomes in the raster grid, defined as observing 16, 36, and 100 outcomes per 10 x 10 area, respectively.

Outcomes are generated using a \textit{spatial additive decay effect} function. Specifically, the observed outcome at location $x$ is:

\begin{equation}
Y_x(\mathbf{Z}) = Y_x(0) + \sum_{i=1}^N \text{Pois} \left(\frac{100}{\max(d_{ix}, \epsilon)}\right)Z_i  \mathbb{I}\{d_{ix} \leq 10\}
\end{equation}

\noindent where $d_{i x}$ is the distance from location $x$ to node $i$, and $\mathbf{Z}$ is the treatment assignment vector. $Y_x(0)$ is the control outcome at point $x$, chosen to be zero for simplicity. The individual effect of node $i$ on location $x$ is generated Poisson with parameter $\left\{\frac{100}{\max(d_{ix}, \epsilon)}\right\}$ where $\epsilon$ is a small positive number. The indicator in the outcome formula ensures that the local interference assumption is satisfied.

\begin{figure}[ht]
    \centering
    \includegraphics[width=0.85\linewidth]{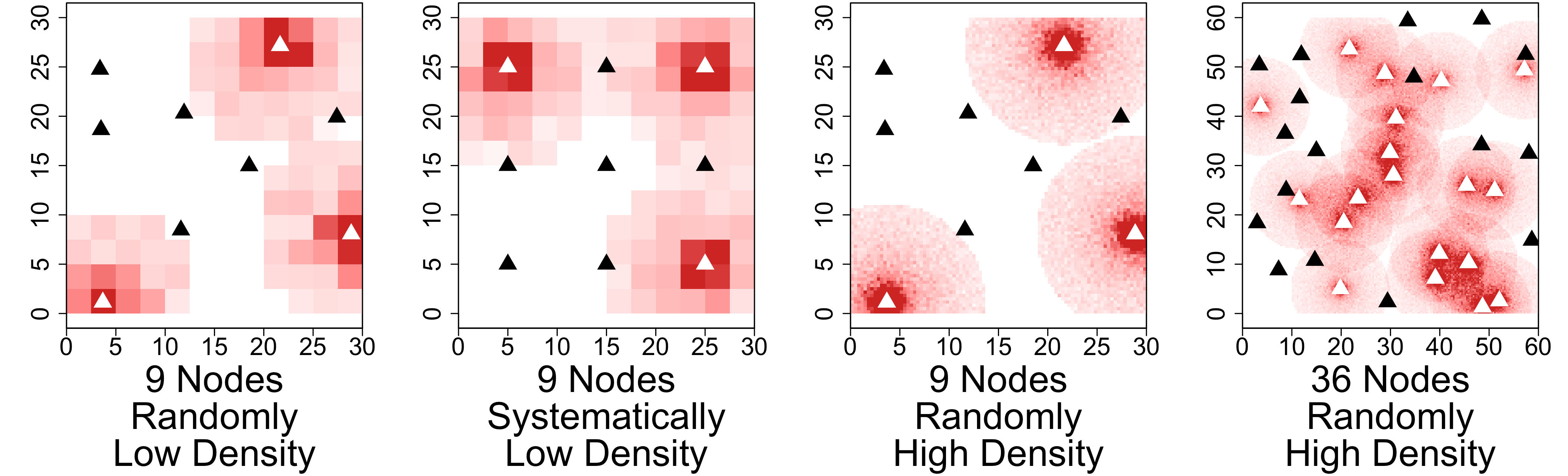}
    \caption{Outcome space for various experimental settings under spatial additive effect process. White triangles represent treated nodes while black triangles represent control nodes. The nodes are overlayed on a heatmap of the outcome raster.}
    \label{fig:sim-examples}
\end{figure}

This decay effect was chosen because previous research has shown that crime increases exponentially with increasing distance from police stations \citep{fondevila2021crime}. Figure \ref{fig:sim-examples} includes examples of different experimental settings and their observed outcomes under this spatial additive decay effect process. In the Appendix \ref{otherdgp}, we include simulations with two additional data generating processes. The first considers a null effect where $AME(d) = 0$, and the second simulates an additive displacement process.

We analyze the performance of the Horvitz-Thompson and H\'ajek estimators for the AME using three spatial models to predict outcomes previously described: the discretized model, kriging, and inverse distance weighting. We benchmark the performance of the estimators against the oracle Horvitz-Thompson and H\'ajek estimators, which assume that we could observe a random subset of points along the circle such that the estimated circle averages are unbiased for the true circle average.

\section{Simulation Results} \label{sec:sim_results}

\begin{figure}
    \centering
    \includegraphics[width=1\linewidth]{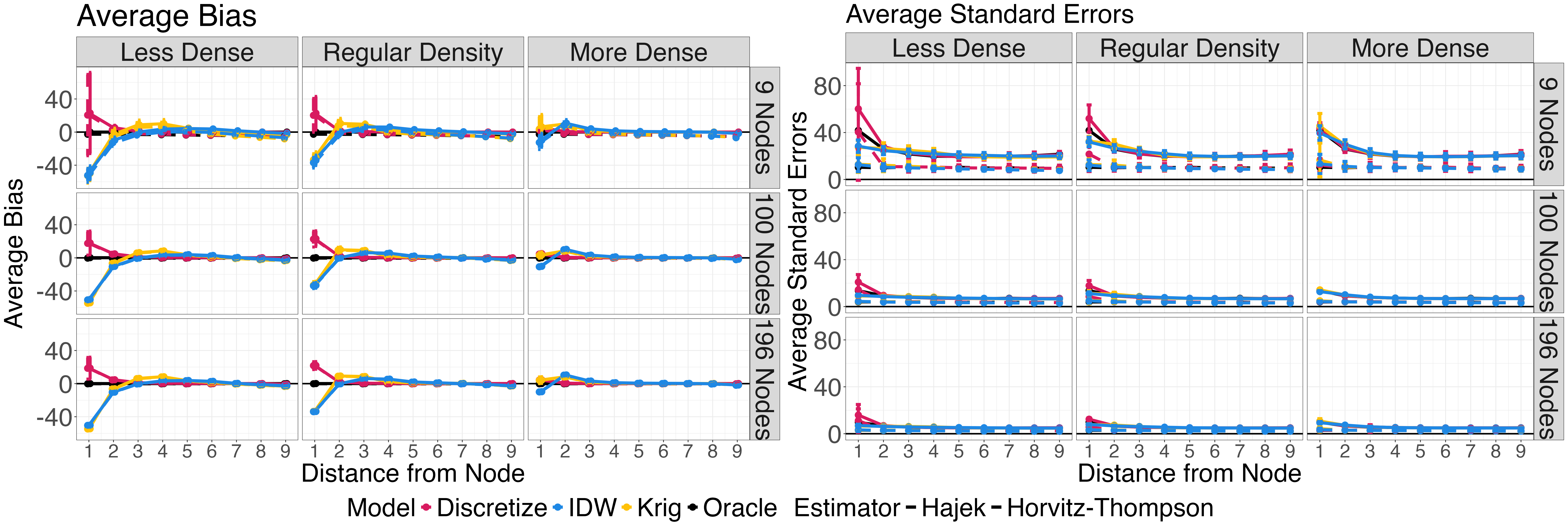}
    \caption{Simulation results with randomly placed nodes. Average bias is across 50 random node placements, and error bars capture variation across node placements. Facet columns vary outcome density, while facet rows vary number of nodes. Three outcome models and two estimators are presented. }
    \label{fig:sim_random}
\end{figure}

We present the bias and standard errors of the Horvitz-Thompson and H\'ajek estimators using different spatial methods for estimating outcomes across a range of distances.\footnote{With $N = 9$ nodes, we calculate the bias and standard errors of the estimators using all of the $2^9 = 512$ possible treatment assignments to the intervention nodes. In the larger $N$ setting, we approximate performance by randomly selecting 500 treatment assignments out of the $2^N$ assignments.} Each plot contains nine subplots that each correspond to outcome density (increasing across columns) and number of intervention nodes (increasing across rows). In Figure \ref{fig:sim_random}, average bias with randomly placed nodes is relative to the average AME across the 50 node placements with error bars capturing variation across placements.

\begin{figure}
    \centering
    \includegraphics[width=\linewidth]{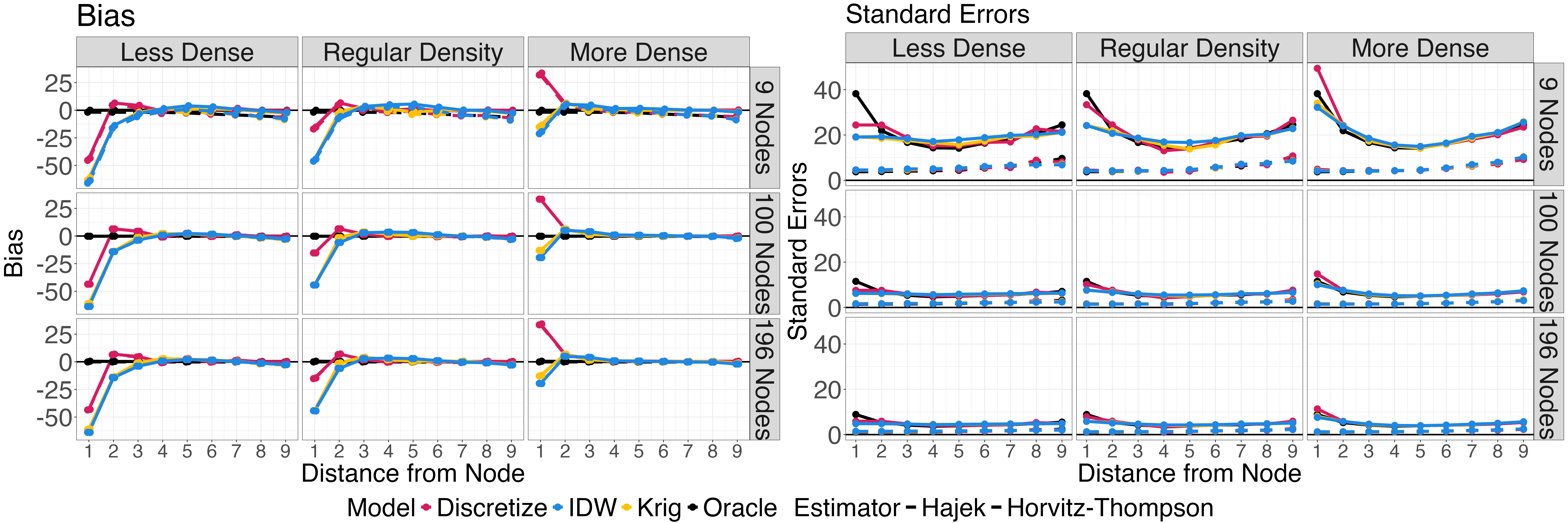}
    \caption{Simulation results with systematically placed nodes. Facet columns vary outcome density, while facet rows vary number of nodes. Three outcome models and two estimators are presented.}
    \label{fig:sim_nr}
\end{figure}

As seen in Figure \ref{fig:sim_random} and \ref{fig:sim_nr}, the Horvitz-Thompson estimator is design-unbiased, and the H\'ajek estimator is consistent, using oracle outcomes in all cases. However, with estimated outcomes, our simulations show that in practice, even the Horvitz-Thompson estimator accrues bias as a result of this outcome modeling.  We find that bias can be particularly large at small distances from intervention nodes.  This is problematic for applications, such as our crime example, where effects may be larger or of most important substantive interest \citep{fondevila2021crime} at smaller distances. Recalling that we use rasterized outcome measurements, for small distances from intervention nodes there are fewer observed outcomes at the desired distance, placing more emphasis on the modeling assumptions. 

When we have randomly located intervention nodes, the average bias of the estimators using the outcome models decreases with increasing outcome density, but not in increasing number of nodes, as illustrated across the columns of Figure \ref{fig:sim_random}.  Again, for low density and small distances, observed rasterized data is more likely to be at larger distances than the target distance regardless of the number of intervention nodes, which can lead to bias in estimation of the AME.

We now turn to the standard error of the AME estimators using the estimated outcomes.  As the density of the observed outcomes increases, the standard errors of the estimators with outcome models converge to the true standard error with the oracle. In Figure \ref{fig:sim_random}, moving from left to right across the columns, the colored curves correspond to the standard errors using estimated outcomes in the AME estimator collapse to the black, oracle curve. As outcome density increases, the outcome model uncertainty decreases, and therefore the standard errors of the AME estimators also decrease. The estimators are consistent in the number of nodes, although the specific convergence rates depend on the convergence rate of the outcome models compared to that of the oracle estimator which is $O(1/\sqrt N)$.

When intervention nodes are systematically placed, we see that the bias incurred from using outcome models (see left panel of Figure \ref{fig:sim_nr}) can be much higher than when nodes are randomly located.  If using the discretized model, the bias may not decay monotonically with increasing density, especially when estimating effects at small distances. This can happen if the observed outcomes in the lower density setting happen to be close to the circles. In Figure \ref{fig:sim_nr}, we see that the standard errors with the discretized model do not converge to the oracle with increasing density. Kriging and IDW suffer less with systematic node locations because these models are smoother. However, with these models, the bias incurred with systematically located nodes is still higher than that with randomly located nodes.

\subsection{Choosing the Best Outcome Model}
We have shown that the choice of outcome model can impact the estimation of the AME. In this section, we briefly discuss how to pick an outcome model.  It is common to use cross-validation to choose among potential estimators. However, that would require that we have observed outcomes, yet our circle distances are estimated using predictions from our model, so we cannot directly choose a model to best fit the circle distances.  Instead, we propose a reasonable alternative in which we use k-fold cross-validation of the three models under consideration using lagged outcomes where fit is measured among the observed outcomes.  We suggest lagged outcomes for cross-validation to avoid overfitting to the treatment effect curve. Using the lagged outcomes is in the spirit of a locked box approach  \citep{rubin2008objective} where post-treatment outcomes are only used in estimation of the AME. For each model class under consideration, we fit the model and generate predictions for observed outcome locations leaving out one fold of the data. We assess model fit using the mean squared error of the model's predictions to the true lagged outcomes in the left out fold. We repeat this procedure for each of the k-folds. Note that we suggest assessing fit using observed outcomes, which are points, rather than the circle averages since these are directly unobservable. Moreover, using lagged outcomes may approximates a null-effect setting, in which many models perform similarly (see Appendix \ref{otherdgp}). Section \ref{sec:discussion} further discusses these problems and suggests alternative approaches, including for when lagged outcomes are unobserved. 

\section{Application}

We return now to our re-analysis of \citet{collazos2021hotspot}.  In their original analysis, \citet{collazos2021hotspot} use estimating equations to estimate direct effects and spillover effects to nearby streets. When accounting for potential short- ($\le$125m) and long-range (125-250m) effects, they find no statistically significant direct or spillover effects. Unlike the approach in \citet{collazos2021hotspot}, the AME framework bypasses the need for estimating equations and allows for arbitrary interference patterns. The interactions between direct, short, and long range spillover effects do not have to be specified. \citet{puelz2022randomization} also reanalyze this data using a graph theoretic approach for randomization tests.  
They find suggestive evidence of spillover effects for nodes between 225m and 425m from intervention nodes; however, when adjusting for known covariates, these effects are no longer statistically significant. 

Our primary observed outcome is a street-level aggregated crime index. This harm index \citep{sherman2016cambridge} is a weighted sum of reported homicides, assaults, motorbike and car thefts, and personal robberies, weighted by the average sentence for each type of crime in Colombia.\footnote{These weights are 0.550 for homicides, 0.112 for assaults, 0.221 for car and motorcycle thefts, and 0.116 for personal robbery.} We also estimate effects for each of these crimes separately. Unlike our simulation setup, outcomes are available for all street segments, which are not arranged on a grid. Figure \ref{fig:medellin-streets } plots the coordinates of all the street segments, treated hot spots, and control hot spots. The density of observed outcomes is on average around 0.0003 observed outcomes per square meters, or about 12-14 street segments within 100m of intervention nodes. 

We estimate the AME at a range of 50 to 1,000 meters in 50-meter increments. This choice is motivated by prior research that suggests longer-range spatial spillovers.
\citet{fondevila2021crime} find that most crime occurs within 1 km of police stations in Argentina. In the same experiment we consider, \citet{puelz2022randomization} find suggestive evidence of spillover effects for nodes between 225 and 425 meters from intervention nodes. Given these findings, we study spillovers beyond the 250 meters initially studied in \citet{collazos2021hotspot}.

We note that these outcomes are reported crimes provided by the police department rather than confirmed crimes. \citet{collazos2021hotspot} discuss how measurement error from the willingness to report could be correlated with treatment. Reporting rates among citizens could increase with treatment if increased policing garners increased trust. They could also decrease with treatment if police patrols are incentivized to not receive reports in areas with intervention \citep{brantingham2021public}. To address this limitation, \citet{collazos2021hotspot} additionally analyze baseline and endline surveys, and these results suggest a major improvement in security perceptions.  These surveys are not analyzed here.

\subsection{Choosing the Best Outcome Model}
As discussed above, we use $k$-fold cross validation to compare three possible outcome modeling approaches: discretized, kriging, and IDW.  Table \ref{table:cv-results} contains the results from k-fold cross-validation with four folds of the three models under consideration using lagged outcomes.  We also considered using a donut estimator as described in \citet{Wang2025aoas}, but this approach was infeasible; see Appendix \ref{donut} for details. The discretized model requires hyperparameter tuning for the dimensions of the raster, i.e. how fine or coarse the outcome space is, which we discuss in Appendix \ref{disc-hyper}. 

We find that IDW outperforms the discretized model and kriging in our application, and select it as our primary outcome model in our analysis. In our data, the spatial autocorrelation is low. We used Moran’s I index to test for the amount of spatial autocorrelation and found that the observed statistic is close to zero and statistically significant, indicating little to no spatial autocorrelation in the data. Kriging incorporates autocorrelation into the model but may not perform better than IDW if no correlation is present. Through cross-validation, we found that IDW performs best across a range of hyperparameters controlling the number of adjacent points used for prediction for both models.

\begin{table}[ht]
\centering
    \begin{tabular}{l c}
        \toprule
        \textbf{Model} & \textbf{Mean Squared Error} \\
        \midrule
        Discretized & 0.0310 \\
        Kriging & 0.0297 \\
        IDW & 0.0269 \\
        \bottomrule
    \end{tabular}

\caption{Results from cross-validation on lagged, pre-treatment outcomes for three models under consideration.}
\label{table:cv-results}
\end{table}

We further use cross-validation to select hyperparameters for IDW (see Appendix \ref{disc-hyper}). We assess the robustness of our results to the other modeling options in Appendix \ref{robustness-model}.

\subsection{Results}

\begin{figure}[ht]
    \centering
    \includegraphics[width=0.75\linewidth]{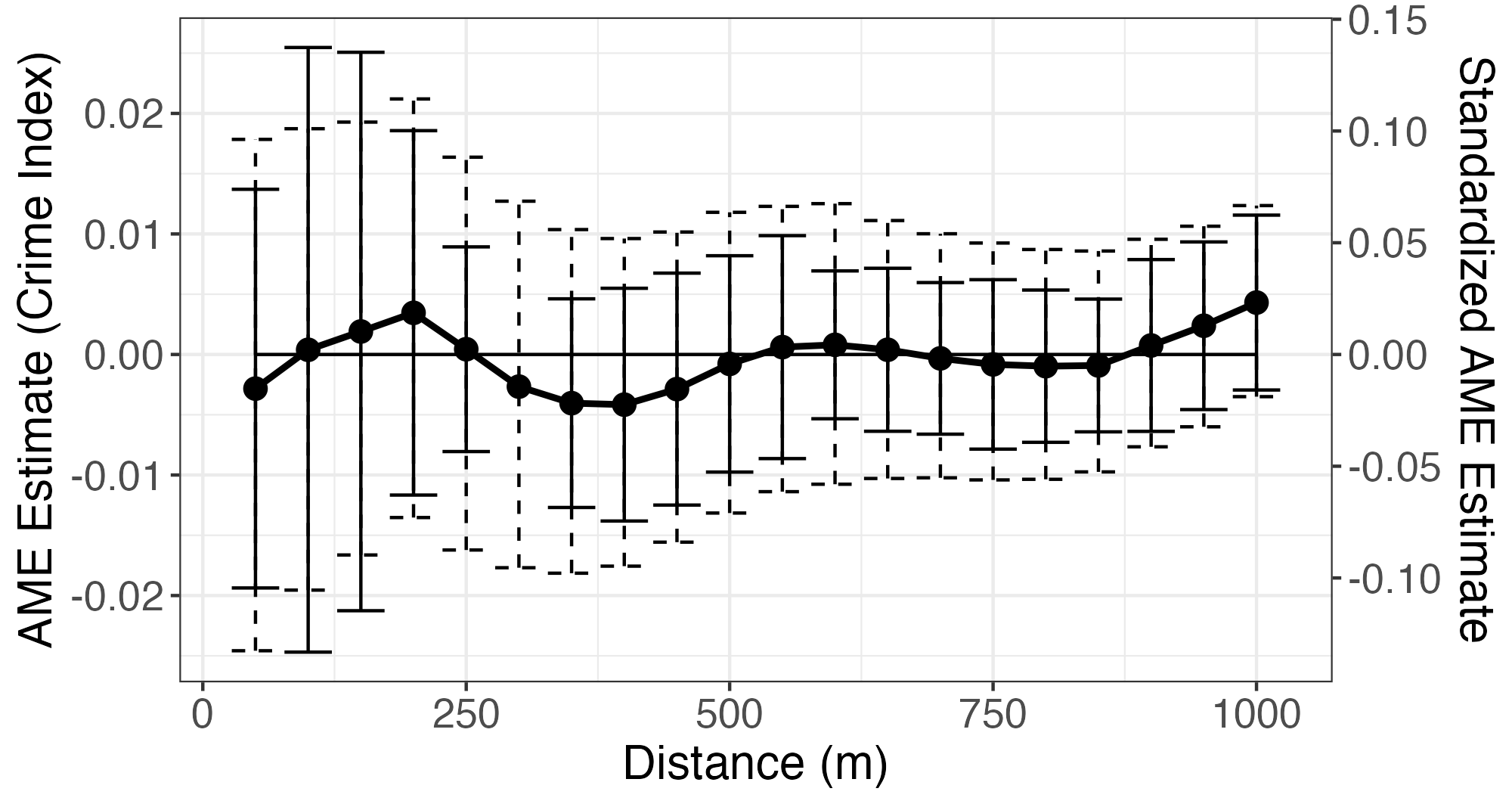}
    \caption{AME estimates for the effect of hot spots policing on crime using first-differenced outcomes. Effects estimated using H\'ajek estimator and IDW for estimating outcomes.  95\% confidence intervals are presented from estimation of the Conley Spatial HAC standard errors (solid) and inversions of permutation tests (dashed).}
    \label{fig:med-final}
\end{figure}

Figure \ref{fig:med-final} presents the H\'ajek AME estimates for the crime index using the IDW model to estimate the circle averages, incorporating the unequal probabilities of treatment. The left axis shows AME estimates on the scale of the index, and the right axis is scaled in terms of standard deviations of the pre-treatment crime index outcome.   The resulting AME curve is low in magnitude and statistically insignificant across all distances.\footnote{We additionally estimate smoothed AME curves as suggested by \citet{Wang2025aoas}. Across different bandwidths for the kernel, the curves remain low in magnitude and statistically insignificant.} These results support the findings from \citet{collazos2021hotspot} which found no significant spillover effects. Consistent with existing literature, our results suggest that hot spots policing does not displace crime to further locations \citep{braga2019hot, braga2014hotspots, weisburd2006does}. Although we cannot rule out small displacement effects, it is unlikely that there are large displacement effects up to 1 kilometer. Appendix \ref{app:d0} contains results for the direct effect when $d=0$. 

We present two types of confidence intervals. First, 95\% confidence intervals are formed using the Conley spatial HAC standard errors as suggested by \citet{Wang2025aoas}.  Second, we present 95\% permutation-based confidence intervals formed by inverting permutation tests of the sharp null.  Unlike Conley standard errors, permutation tests do not rely on model assumptions and are often preferred by experimentalists for their design-based inference.  Under interference, \citet{rosenbaum2007interference} shows that standard permutation test inversion forms a $100(1 - \alpha)$\% confidence interval for a slightly modified estimand, $\Delta$, where $\Delta$ is the difference in the magnitude between the AME statistic under the given experiment and a placebo experiment (i.e. one with a sharp null of no effect). While the interpretation of the null slightly differs from those formed with Conley standard errors, the advantage is the intervals are exact under the design. Our permutation based intervals account for the fact that each hot spot has an unequal probability of treatment, approximated by an unequal-probability Bernoulli design.\footnote{\citet{collazos2021hotspot} randomized treatment according to restrictions from the Metropolitan police which we are unable to replicate for our permutation testing. However, using a Bernoulli design with the given probabilities of treatment is likely a sufficient approximation to the true randomization scheme.}

\begin{figure}[h]
     \centering
     \begin{subfigure}[b]{0.5\textwidth}
         \includegraphics[width=\textwidth]{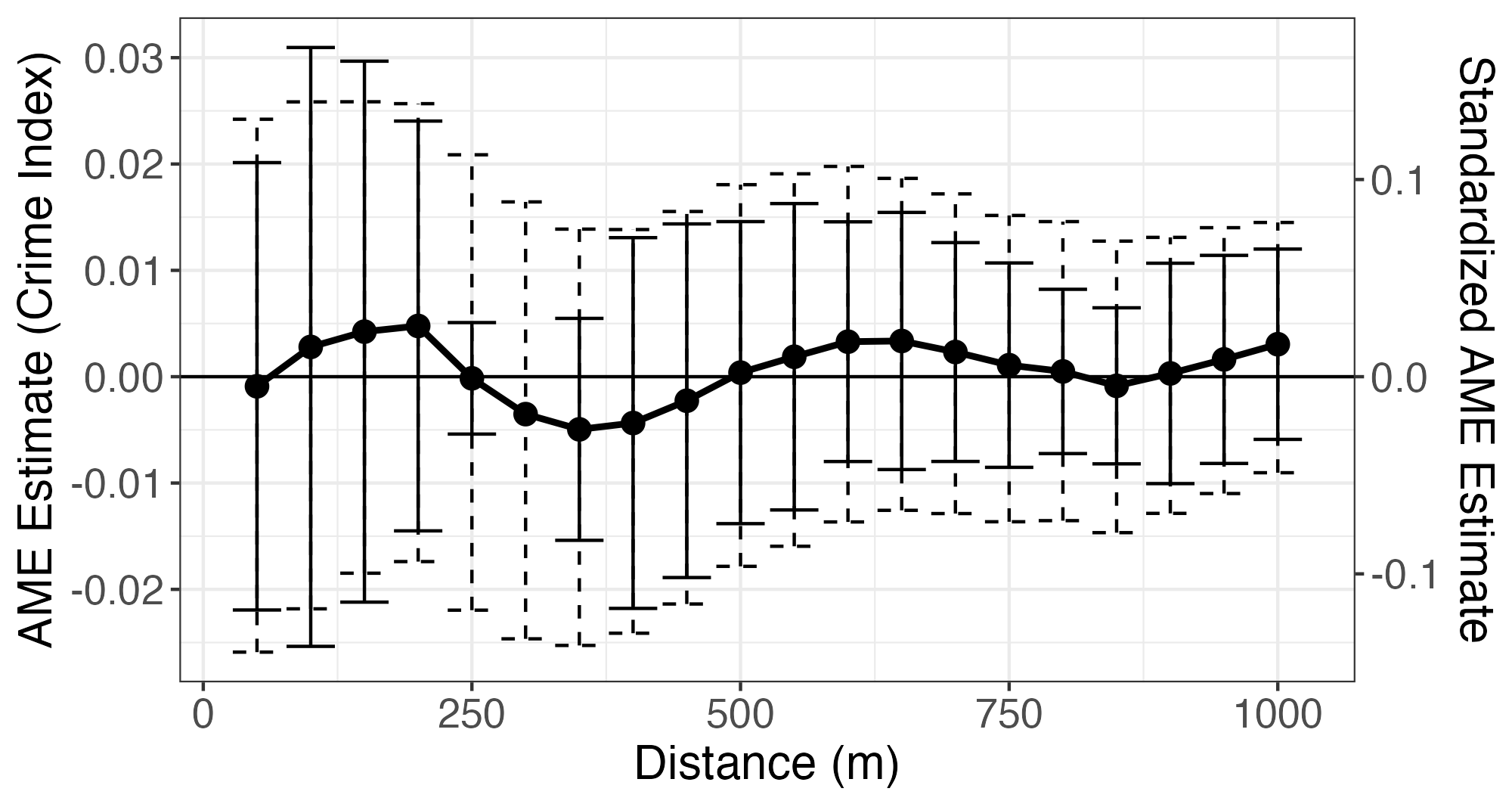}
         \caption{Pre-Treatment (Baseline) Outcomes}
         \label{fig:med_lagged} 
     \end{subfigure}
     \begin{subfigure}[b]{0.5\textwidth}
         \includegraphics[width=\textwidth]{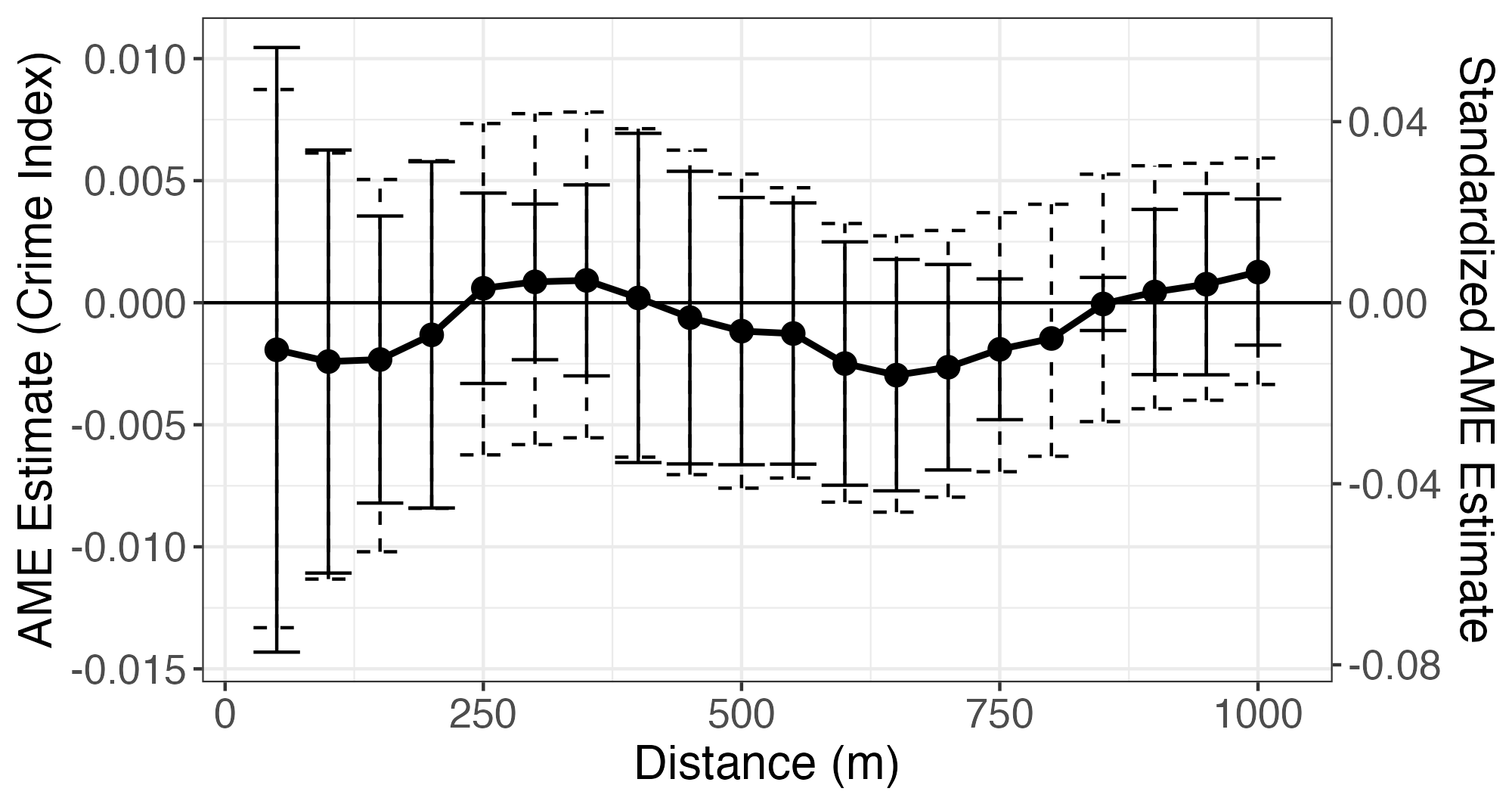}
         \caption{First-differenced Outcomes}
         \label{fig:med-fd} 
     \end{subfigure}
     \caption{AME estimates for the effect of hot spots policing on crime as measured by the crime index. Effects estimated using H\'ajek estimator and IDW for estimating outcomes.  95\% confidence intervals are presented from estimation of the Conley Spatial HAC standard errors (solid) and inversions of permutation tests (dashed).}
\end{figure}

It is common to test for covariate imbalance to bolster the claim that the estimate is unlikely to be an outlier.  While the original authors present node-level balance statistics, the advantage of the AME is that we can also conduct a placebo, or negative control, analysis using lagged outcomes \citep{2015causal}.  Figure \ref{fig:med_lagged} shows the AME curve of the lagged crime index outcome from 2014 using our selected IDW model.  The AME curve for the lagged outcome is also low in magnitude and insignificant across all distances, indicating balance on prognostic outcomes. 

However, since the point estimates are not all zero, we also analyze the first-difference from the lagged outcomes, which can account for any baseline differences in the crime index between streets. Figure \ref{fig:med-fd} shows the AME curve for the first difference of crime index.  Effects are also substantively small and statistically insignificant, strengthening the result that hot spots policing may not reduce or displace crime.

\begin{figure}[ht]
    \centering
    \includegraphics[width=1\linewidth]{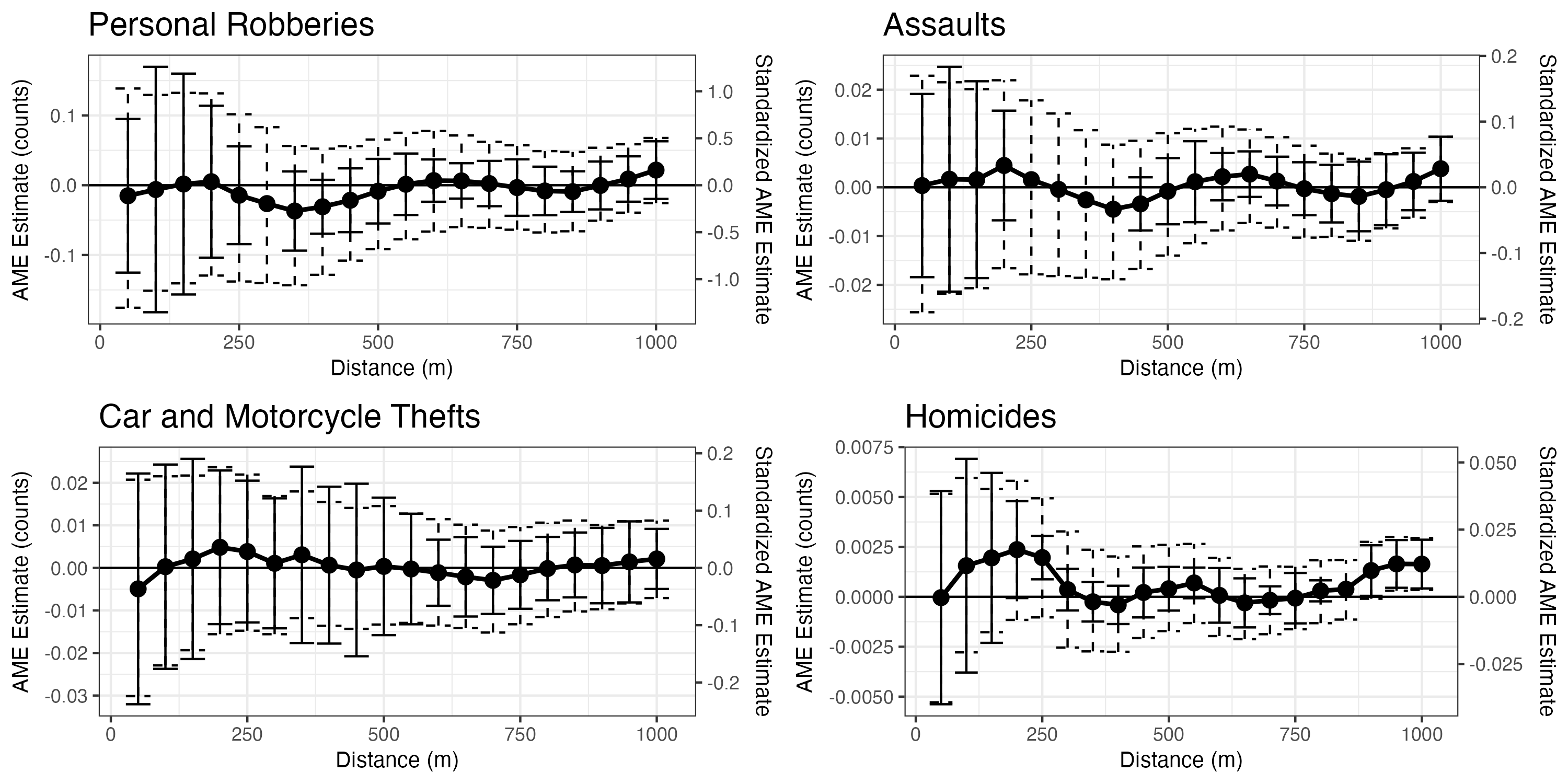}
    \caption{AME estimates for the effect of hot spots policing on individual crime counts. Effects estimated using H\'ajek estimator and IDW for estimating outcomes.  95\% confidence intervals are presented from estimation of the Conley Spatial HAC standard errors (solid) and inversions of permutation tests (dashed).}
    \label{fig:sep_outcomes}
\end{figure}

Finally, while the overall crime index shows no statistically significant effects, we wish to  understand if there are differences in effects among crime types. Figure \ref{fig:sep_outcomes} includes the disaggregated results for each of the four crime outcomes that comprise the crime index.  We see null effects for all crime outcomes. This result agrees with \citet{collazos2021hotspot} which finds that assuming short and long range spillovers, hot spots policing had no significant effect on any of the four types of crime considered.

\section{Suggestions for Practice} 
Our simulations and analysis motivate several suggestions for practice when using design-based methods for spatial interference in both the design stage and analysis stage. In the design stage, if researchers have control over where the intervention nodes will be located, they should place the nodes randomly in space if they plan to observe outcomes on a grid. Our simulations show that having intervention nodes systematically located relative to the observed outcomes can cause considerable bias. 

In many cases, researchers are not able to control the location of the intervention nodes, but may have control over where outcomes are measured. For example, in estimating the effect of policing on crime, the location of the street blocks that can be intervened on are often fixed and systematically arranged in a grid.  In these cases, where researchers have control, we do not recommend observing outcomes on a grid, and instead, recommend observing outcomes at random points in space. 

When researchers are interested in estimating effects at specific distances from intervention nodes, our second design-stage suggestion is to measure outcomes at random points along circles with those specific radii. For example, if a researcher is interested in distances of 1, 2, and 3 kilometers, observing outcomes along circles of these radii away from  nodes improves estimation at those distances by bypassing the need for outcome modeling. This might be more important, in particular, for small distances and under a tight data collection budget. However, improved estimation at specific distances may come at the cost of potentially poor estimation of the AME at other distances.  Related, our simulations have shown the impact of outcome density on the performance of design-based spatial estimators.  If possible, researchers should observe more outcomes at small distances from intervention nodes. Alternatively, researchers can modify the estimand by switching the shape $\Omega_d$ to a donut or disk when few outcomes are observed.

In addition to design-stage recommendations, our results motivate suggestions for practice in the analysis stage. In the analysis stage, the main decision researchers must make is what model to use. For example, our simulations point to situations where the discretized model may cause considerable bias. If nodes are systematically located in relation to observed outcomes, researchers should avoid using the discretized model. Even if the density of observed outcomes is high, the discretized model can still result in high bias.  For studies with few intervention nodes as well as few observed outcomes, researchers should also avoid using the discretized model.  At small distances, IDW or kriging may be better because the smoothing results in less variable estimates. In our application, we illustrate how cross-validation can be used to select among candidate outcome models.

\section{Discussion}\label{sec:discussion}
Even under design-based frameworks for spatial settings, outcome modeling is necessary to estimate spillover effects in many spatial settings. Both existing design-based frameworks, \citet{Wang2025aoas} and \citet{pollmann2023causalinferencespatialtreatments}, seek to estimate spillover effects through a contrast between average outcomes within a shape (circle, donut, or disk) a distance $d$ away from treated locations to average outcomes within a shape a distance $d$ away from control locations. These circle, donut, or disk average outcomes often cannot be directly observed; instead, only outcomes at specific points in space, often arranged in a grid, are observed. Therefore, applied researchers must estimate circle averages through outcome modeling. Even the Horvitz-Thompson estimator of the AME can accrue bias as a result of the outcome modeling. The density of the observed outcomes impacts the performance of the estimators using outcome modeling. 

In both our simulations and data application, we have shown that the choice of outcome model matters when estimating effects. This modeling choice can have a particularly large impact on the performance of the estimators at small distances, where effects are often most important and likely to occur. We have suggested some design-stage solutions to mitigate the estimation error. We have also suggested several ways of thinking about how to choose the best model for a particular setting using cross-validation.

Our work relates to the broader spatial causal inference literature. While we focus on design-based inference in settings where the number of intervention locations is finite  \citep{Wang2025aoas, pollmann2023causalinferencespatialtreatments},  \citet{papadogeorgou2022spatio} works in settings where treatment and outcome variables are generated by spatio-temporal point processes. Their proposed estimator also uses spatial smoothing of the outcome point process. \citet{borusyak2023shocks} propose a design-based approach for estimating treatment effects that combine multiple sources of variation. Though general, their regression based approach can be used for estimating spillovers from randomized interventions. \citet{leung2022design} studies cluster-randomized designs in which spatial regions are randomized into treatment, rather than specific locations as in the present paper. Moreover, we focus on frameworks with arbitrary interference patterns while \citet{leung2022design} defines a model for spatial interference. We also note that there is a growing literature on approaches to address spatial confounding \citep{papadogeorgou2019adjusting, papadogeorgou2023spatial, gilbert2021causal, woodward2025understanding, pollmann2023causalinferencespatialtreatments}. Future research is needed to understand whether findings from our paper translate to these settings. Lastly, there are also existing model-based approaches that deal with spatial confounding \citep[e.g.][]{paciorek2010importance}.

Our work also relates to a small, but growing literature on measurement errors in outcomes for causal inference. With remote sensing or satellite image data, machine learning is often used to generate predictions for outcomes at locations not directly observed. \citet{gordon2024remotesensing} show how machine learning measurement errors can bias causal inference estimates if these non-random errors are correlated with treatment variables. Like our work, they argue that these measurement errors can bias estimates even in experimental or quasi-experimental settings. They propose the use of an adversarial debiasing algorithm for bias correction. Similarly, \citet{shu2019measurement} demonstrate that inverse probability weighting estimators can incur bias with measurement errors for continuous outcomes, unless the error model is an additive function of the true outcome. Outcomes annotated by large language models can also suffer from measurement errors, and \citet{egami2024using} develop a framework to correct this issue using a smaller set of accurately measured outcomes.

This spatial setting parallels clustered randomized trials. Points that lie within the shape of interest (e.g. circles) can be thought of as being part of the same cluster, and within each cluster, there is a sample of individual outcomes that contribute to the cluster-level outcome. Previous work on clustered-randomized trials has studied the precision tradeoff between the number of clusters and the number units within each cluster. However, the present setting differs slightly from these settings because the units within clusters may not be independent due to the spatial interference. One point's outcome can contribute to the circle average of two different nodes.

Future research is needed on the impact of outcome modeling on uncertainty estimation in these design-based frameworks for spatial spillover.  We present Conley Spatial HAC estimators suggested by \citet{Wang2025aoas} as well as permutation testing for inference.  However, neither Conley spatial HAC estimators or these permutation tests account for uncertainty from modeling of the outcomes. The permutation tests are valid given a specific outcome model; however, they may not provide the same inferential results as permutation tests conducted with an oracle estimator.   In accordance with \citet{gordon2024remotesensing}, we believe further research is needed on how to correct standard errors to account for the uncertainty in estimation of the circle averages. 

A fruitful avenue for further research is also in the area of model selection. The necessity of using outcome models for design-based spatial causal inference methods begs the question of how to pick the model in a principled way.  Model performance on observed point outcomes can be used as a proxy, but it does not directly target model performance when estimating circle averages. In our application, we use cross-validation on experimental and lagged outcomes to assess model performance. Although lagged outcomes are in the spirit of \citet{rubin2008objective} locked box approach, our simulations show that many outcome models may perform well on null effect processes (i.e. lagged outcomes) but poorly on spatial additive decay effect processes (i.e. experimental outcomes). Moreover, lagged outcomes may be unobserved, requiring an alternative strategy for model selection. Spatial interference complicates existing sample splitting frameworks like honest inference \citep{athey2016recursive} because splits may no longer be independent. A simple workaround could be to split the space into two disjoint regions sufficiently separated by a buffer zone. When the two regions are far enough apart as to be unaffected by interference, one region would be used for outcome model selection while the other would be left for estimation. Future research is needed to more formally explore such approaches for model selection under spatial interference. 

We have highlighted the importance, and often necessity, of outcome modeling in experimental design-based inference in spatial settings.  We show how common estimators for outcomes can lead to bias in estimation of causal effects, even for the Horvitz-Thompson estimator which is design-unbiased with oracle outcomes.  Through a novel analysis of \citet{collazos2021hotspot}, we show how analyzing the AME across distance can provide a more nuanced understanding of the effect of hot spots policing.  We hope that the guidance and illustrative example here can help applied researchers in analyzing spatial spillover effects.

\newpage
\printbibliography[
heading=bibintoc,
title={References}
] 

\pagenumbering{arabic}
\renewcommand*{\thepage}{A\arabic{page}}
\setcounter{section}{0}
\renewcommand*{\thesection}{A\arabic{section}}
\setcounter{figure}{0}
\renewcommand*{\thefigure}{A\arabic{figure}}
\setcounter{table}{0}
\renewcommand*{\thetable}{A\arabic{table}}

\newpage
\appendix

\begin{center}
\Large
\textbf{Appendix}
\end{center}

\label{sec:appendix}
\begin{refsection}
\onehalfspacing

\section{Setup}
\subsection{Local interference assumption} \label{sec:local-interference}
 We follow \citet{Wang2025aoas} in describing the local interference assumption. Let the indicator $I_{ij}(d)$ denote whether assignment at intervention node $j$ interferes with the $d$-radius circle average at node $i$. This variable takes on a value of one if changing the treatment assignment of $j$ changes the value of node $i$'s circle average. Let the indicator $s_{ij}(d)$ denote whether node $i$ and $j$'s circle averages both depend on some other node, $l$. $l$ could be $i$, $j$, or some other node. This variable takes on a value of one if both $I_{il}(d)$ and $I_{jl}(d)$ are one. In this case, changing treatment assignment at $l$ changes the circle averages at both $i$ and $j$. Let $d_{ij}$ be the distance between nodes $i$ and $j$.

The local interference assumption is as follows: Let $h: [0, \infty) \xrightarrow{} [0, \infty)$ be a function independent of sample sizes. For each $d$, and all pairs of intervention nodes $i$ and $j$, there exists a constant $h(d)$ such that if $d_{ij} > h(d)$, then $s_{ij}(d) = 0$. 

This assumption means that nodes beyond some distance are not dependent on each other nor do they share any sources of variation. \citet{Wang2025aoas} provide an interpretable sufficient condition for this assumption. Assume that intervention at every node does not influence outcome points that are located $\bar d$ or farther away from the node. The local interference assumption is satisfied if $h(d) = 2\bar d + 2d$. 

\section{Analytic Results under Discretized Model}\label{app:theory}

\subsection{Discretized Model and Circle Average}
As discussed in the main text, we consider the estimation error a single circle average of radius $d$ centered at intervention node $c \in \mathbb{R}^2$. 
The outcome of a location $x$ is fixed and an attribute of the location, and $Y(x) = f(x) + \epsilon_x$. Recall that without loss of generality, $f(x)$ is only a function of the radial distance to the node $c$. 

To formalize the difference between randomly and systematically placed nodes, we let the intervention node center $c$ be a random variable. For systematically placed nodes, the center takes on only one position with probability 1. This setup allows us to consider both cases in one mathematical argument.

\subsubsection{True Circle Average}
Let $r(\theta, R) = \begin{pmatrix} R\cos(\theta) \\ R\sin(\theta) \end{pmatrix}$. Note that $p = c + r(\theta, R)$, which is the polar representation of $p$.  

For a circle centered at $c$ with radius $R$, the true circle average $\mu(c, R)$ is:
\begin{align}
    \mu(c, R) &= \frac{1}{2\pi} \int_0^{2\pi} Y(c + r(\theta, R)) \, d\theta \\
    &= \frac{1}{2\pi} \int_0^{2\pi} \left[ f(c + r(\theta, R)) + \epsilon(c + r(\theta, R)) \right] \, d\theta\\
    &=  \frac{1}{2\pi} \int_0^{2\pi}  f(c + r(\theta, R)) \, d\theta +  \frac{1}{2\pi} \int_0^{2\pi}  \epsilon(c + r(\theta, R))  \, d\theta\\
    &=  \frac{1}{2\pi} \int_0^{2\pi}  f(c + r(\theta, R)) \, d\theta \quad \text{(mean-zero errors)}
\end{align}

\subsubsection{Estimated Circle Average}
As in the main text, we set up the discretized model by defining a grid in $\mathbb{R}^2$  (which we treat as extending without bound) with each cell being a square of length $L$. Each cell $k$ has a center or midpoint with coordinates $g_k \in \mathbb{R}^2$. 
Under the discretized model, the value at any point $p$ is modeled as the value of the grid it falls in. A grid takes the value of the center of the grid, $g_k$. The discretized model, $D(p)$, is then: 

\begin{equation}
    D(p) := Y(g_k) = f(g_k) + \epsilon(g_k) \quad \text{for } p \in \text{Cell}_k
\end{equation}
A point $p$ on the circle is offset from the nearest grid center by an offset $\delta$:
\begin{equation}
    g_k = p + \delta \quad \text{with } \delta \in \left[-\frac{L}{2}, \frac{L}{2}\right] \times \left[-\frac{L}{2}, \frac{L}{2}\right]
\end{equation}

The magnitude of the offset can be at most $\frac{L}{2}$ in both the $x$ and $y$ direction since the cell is of size $L$. We use $\delta_x$ and $\delta_y$ to denote the $x$ and $y$ components of the offset.  

\begin{figure}[h]
    \centering
    \includegraphics[width=0.75\linewidth]{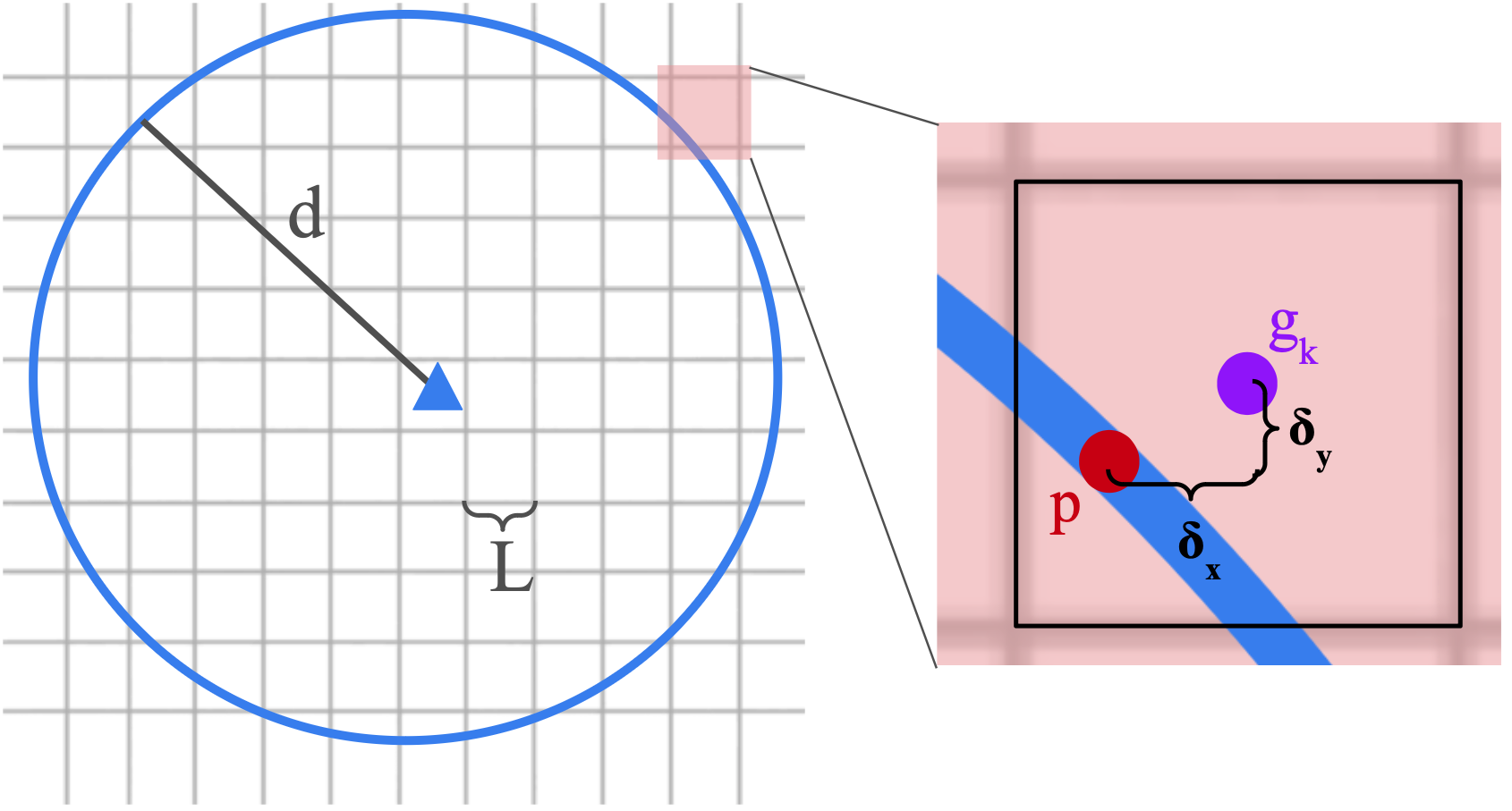}
    \caption{Diagram showing overall setup, point of interest $p$, center of cell $g_k$, and offset. }
\end{figure}

The circle average estimator using the discretized model is:
\begin{equation}
    \hat{\mu}(c, d) = \frac{1}{2\pi} \int_0^{2\pi} D(c + r(\theta, d)) \, d\theta
\end{equation}

Our goal is to characterize the bias of this estimator. 

\subsubsection{Bias in Circle Average}
Let $e(p)$ denote the error between the modeled outcome and the true outcome at a point $p$: 
$$e(p) := D(p) - Y(p)$$

Then, the bias in the circle average across node placements is:

\begin{align}
    \mathbb{E}_c[\hat{\mu}(c, d)] - \mu(c, d) 
    &= \mathbb{E}_c\left[ \frac{1}{2\pi} \int_0^{2\pi} D(c + r(\theta, d)) \, d\theta\right] - \frac{1}{2\pi} \int_0^{2\pi} Y(c + r(\theta, d)) \, d\theta\\
    &=  \mathbb{E}_c\left[ \frac{1}{2\pi} \int_0^{2\pi} D(c + r(\theta, d)) - Y(c + r(\theta, d)) \, d\theta\right]\\
    &= \mathbb{E}_c \left[ \frac{1}{2\pi} \int_0^{2\pi} e(c + r(\theta, d)) \, d\theta \right]\\
    &= \frac{1}{2\pi} \int_0^{2\pi} \mathbb{E}_c[e(c + r(\theta, d))] \, d\theta 
\end{align}

The last line is from Fubini's theorem. It shows we can focus on the expression inside the integral: the expected error for a particular angle, $\theta$, across the random variable, $c$.

\subsection{Taylor Expansion at a Single Angle}

Recall that the discretized model prediction at a point $p$ is $D(p) = f(g_k) + \epsilon(g_k) = f(p + \delta) + \epsilon(g_k)$.  

The error between the modeled outcome and the true outcome is:
\begin{align}
        e_c(p) &:= D(p) - Y(p)\\
        &= \left[f(p+ \delta) + \epsilon(g_k) \right]- \left[ f(p) + \epsilon(p)\right]\\
        &= \left[ f(p + \delta) - f(p) \right] + \left[ \epsilon(g_k) - \epsilon(p) \right]
\end{align}

Above, we include a subscript $c$ for on the error to highlight that the error depends on $f$, which is determined by the location of the node location, $c$.

To characterize the first term, we perform a Taylor expansion of $f$ around $p$:
\begin{align}
        f(p + \delta) &= f(p) + \nabla f(p) \cdot \delta + \frac{1}{2} \delta^T H_c(p) \delta + O(\delta^3)\\
        &\approx f(p) + \nabla f(p) \cdot \delta + \frac{1}{2} \delta^T H_c(p) \delta
\end{align}

where $H_c(p)$ is the Hessian matrix of $f$ evaluated at $p$. 

Therefore:
\begin{align}
    e_c(p) &= \left[ f(p + \delta) - f(p) \right] + \left[ \epsilon(g_k) - \epsilon(p) \right]  \\
    &\approx \nabla f(p) \cdot \delta + \frac{1}{2} \delta^T H_c(p) \delta + \left[ \epsilon(g_k) - \epsilon(p) \right]   \quad 
    \text{(plug in Taylor expansion)}\\
    &= \left( \frac{\partial f}{\partial x} \delta_x + \frac{\partial f}{\partial y} \delta_y \right) + \frac{1}{2} \left( \frac{\partial^2 f}{\partial x^2} \delta_x^2 + \frac{\partial^2 f}{\partial y^2} \delta_y^2 + 2\frac{\partial^2 f}{\partial x \partial y} \delta_x \delta_y \right) + \left[ \epsilon(g_k) - \epsilon(p) \right]
\end{align}

where the last line expands the gradient and Hessian terms.

\subsection{Bias for Systematically vs. Randomly Placed Nodes}
We now look at the difference in bias between systematically versus randomly placed nodes by focusing on the term inside the integral, $ \mathbb{E}_c[e_c(c + r(\theta, d))]$.  Again, this is still for a single angle, $\theta$. We look at the expected error across the random variable, $c$.

\subsubsection{Systematic} \label{app:sys_theory}

With systematic placement, recall there is no randomness in the center. For fixed $\theta$, the point $p = c + r(\theta, d)$ is fixed. If by coincidence, the point lies exactly in the middle of the cell (i.e. $p = g_k$ for the cell it lies in), there is no error and $e_c(p) =0$. Otherwise, there is error:
\begin{align}
        e_c(p) &\approx \nabla f(p) \cdot \delta + \frac{1}{2} \delta^T H_c(p) \delta + (\epsilon(g_k) - \epsilon(p)) \\
        &= O(\delta) +\underbrace{ \epsilon(g_k) - \epsilon(p)}_{\text{error difference}}
\end{align}

The offset is a function of the point $p$. Since $\delta$ scales with $L$, the bias with systematically placed nodes is $O(L)$ plus the exact noise difference at that specific grid point.  

\subsubsection{Random} \label{app:rand_theory}

The center is a random variable uniformly distributed over a grid cell: $c \sim U([0, L]^2)$. Even if we want to randomize beyond that to a larger space, because of symmetry and the repetition of the grid, it suffices to look just at this range. 

We look at the inside of the integral. For any fixed $\theta$, we evaluate at $p = c + r(\theta, d)$. Since $c$ is uniform over a cell of length $L$, $p \sim U([r(\theta,d),  L + r(\theta, d)]^2)$. 

Since $p$ is random, the offset $\delta$ is also random. To find the distribution of the offset $\delta$, we note that it is periodic. As a point moves across $[0,L]$, the offset ranges from $[-L/2, L/2]$. As the point moves across $[L,2L]$, the offset also ranges from $[-L/2, L/2]$. So it suffices to look at the distribution of $\delta$ induced by the distribution of $p \text{ mod } L \sim U([0, L]^2)$. Noting that $\delta_i(p) = L/2 - (p \text{ mod } L)$, we get that: 
$\delta \sim U[-L/2, L/2] \times U[-L/2, L/2]$.

Focusing on one fixed angle:
\begin{align}\label{expect_error}
    \mathbb{E}_c[e_c(c + r(\theta, d))] &= \mathbb{E}_c \left[ \nabla f(p) \cdot \delta + \frac{1}{2} \delta^T H_c(p) \delta + O(\delta^3) + \underbrace{ (\epsilon(g_k) - \epsilon(p))}_{\text{error difference}} \right]\\
\end{align}

We will now analyze each term in the expectation in turn. 

\paragraph{First Term}

The first term is $\mathbb{E}_{c} \left[ \nabla f(p) \cdot \delta \right]$. Recall that $f(p)$ only depends on the distance from the point $p$ to the center $c$. Because $p$ lies on the circle, the distance $||p-c||_2$ is exactly $d$. Even as $c$ moves, for a fixed radius $d$ and angle $\theta$, the gradient $\nabla f(p)$ is a fixed. That is, $\nabla f(p)$ is constant with respect to the random variable $c$. Therefore, we can pull it completely out of the expectation:

\begin{align}
    \mathbb{E}_{c} \left[ \nabla f(p) \cdot \delta \right] &= \nabla f(c + r(\theta, d)) \cdot \mathbb{E}_{c}[\delta] \\
    &= \nabla f(c + r(\theta, d)) \cdot 0 \quad (\delta \text{ mean zero}) \\
    &= 0
\end{align}

The first-order term evaluates to zero due to the uniform distribution of $\delta$. Intuitively, a point is equally likely to have positive offset and negative offset so these cancel out in expectation. 

\paragraph{Second Term}
Now we move to the second term in Equation (\ref{expect_error}), $ \mathbb{E}_\delta \left[ \frac{1}{2} \delta^T H_c(p) \delta \right]$ and use the same logic. Since $f$ only depends on the distance from the point to the center, the Hessian evaluated strictly on the circle boundary is again independent of $c$ for a fixed $\theta$ and $d$. Thus, we can pass the expectation directly to the offset components.

Because the x and y coordinates of the circle are independent, $\delta_x$ and $\delta_y$ are independent, and $\mathbb{E}[\delta_x \delta_y] = \mathbb{E}[\delta_x] \mathbb{E}[\delta_y] = 0$.
Also, $\mathbb{E}[\delta_x^2] = \mathbb{E}[\delta_y^2] = L^2/12$ since each are $U[(-L/2, L/2)]$.

\begin{align}
    \mathbb{E}_\delta \left[ \frac{1}{2} \delta^T H_c(p) \delta \right] &= \frac{1}{2} \mathbb{E}_\delta \left[  \left( \frac{\partial^2 f(p)}{\partial x^2} \delta_x^2 + \frac{\partial^2 f(p)}{\partial y^2} \delta_y^2 + 2\frac{\partial^2 f(p)}{\partial x \partial y} \delta_x \delta_y \right) \right]   \quad \text{ (expand out)} \\
    &= \frac{1}{2} \left ( \frac{\partial^2 f(p)}{\partial x^2} \mathbb{E}_\delta [\delta_x^2] + \frac{\partial^2 f(p)}{\partial y^2} \mathbb{E}_\delta[\delta_y^2] + 2\frac{\partial^2 f(p)}{\partial x \partial y} \mathbb{E}_\delta[\delta_x \delta_y] \right) \\
    &=  \frac{1}{2}  \left( \frac{\partial^2 f(p)}{\partial x^2} \frac{L^2}{12} + \frac{\partial^2 f(p)}{\partial y^2} \frac{L^2}{12} + 0  \right)\\
    &= \frac{L^2}{24} \nabla^2 f(p)
\end{align}

$\nabla^2 f$ denotes the Laplacian of $f$, which is the trace of the Hessian. 

\paragraph{Third Term}

The third term is the expectation over the difference in noise terms: $\mathbb{E}_c[\epsilon(g_k) - \epsilon(p)]$. Since the errors are i.i.d. and mean-zero, $\mathbb{E}_c[ \epsilon(p)] = 0$. That is, as $c$ moves across infinitely many points, $p$ also does and the average of errors is zero. As $c$ moves, $p$ lands in one of at most four cells. For this reason, the law of large numbers does not hold, and $\mathbb{E}_c[\epsilon(g_k)] \neq 0$. 

\paragraph{Overall}

Together, this shows that at a specific angle, estimation error for randomly placed nodes is $O(L^2)$ plus a noise term.  

\subsection{Overall Bias: Assumptions and Comparison}

The previous results focus on a specific angle. To characterize the bias in the circle averages, we integrate over all angles. Below we spell out the needed assumptions and compare results from random and systematic placements. 

\subsubsection{Assumptions}

Our main results hinge on the following assumptions:

\begin{enumerate}
    \item $\nabla f$ and $Hc$ are bounded along the circle. 
    \item Grid is fine relative to the function at radius $d$: 
    $$L \ll \frac{||H_c(p)||}{||f'''(p)||}$$  This is needed for higher-order terms in the Taylor expansion to disappear (be negligible compared to $L^2$ term). 
    \begin{itemize}
        \item To see why, for the approximation to hold we need the $O(\delta^3)$ term to be negligible compared to the $O(\delta^2)$ term. In terms of norms, we need 
        $$\frac{|f'''(p)\delta^3|}{|H_c(p)\delta^2|}   \ll 1$$
        which becomes
        $$\frac{||f'''(p)|| ||\delta||^3}{||H_c(p)|| \delta||^2} \ll 1$$ 
        \item This rearranges to $||\delta|| \ll \frac{||H_c(p)||}{||f'''(p)||}$. 
        \item Since $||\delta||$ scales with $L$, this is the equivalent to assuming $L$ is small relative to the norms. 
        \item Without loss of generality, we assumed that $f$ is a function of the radial distance. If $f$ decays with $d$, this assumption is like assuming that $L \ll d$. For example, if $f(d) = 1/d$ where $d$ is the distance to the center, the ratio of the Hessian to the third derivative is $d$, so $L \ll d$. 
    \end{itemize}
    \item We assume the residual noise term, $\Delta \epsilon$, is negligible compared to the discretization bias from the true effect. Specifically, we assume $\Delta \epsilon = o(L^2)$, allowing the $O(L^2)$ discretization error of $f$ to dominate.
    \begin{itemize}
        \item This occurs when the magnitude of the noise is small relative to the true $f$. 
        \item It also holds when $L$ is small relative to $d$. $\Delta \epsilon$ is the average of $\epsilon(g_k) - \epsilon(p)$ across $\theta$. Recall that the errors are mean-zero and independent. Since we average over $\theta$, $ \frac{1}{2\pi}\int_0^{2\pi}\epsilon(p)d\theta \approx 0$, leaving the term $\frac{1}{2\pi}\int_0^{2\pi}\epsilon(g_k)d\theta$. If $L$ is small relative to $d$, the circle passes through many more distinct cells. By the law of large numbers, $\frac{1}{2\pi}\int_0^{2\pi}\epsilon(g_k) \approx 0$, rendering the residual noise $\Delta\epsilon$ negligible. 
    \end{itemize}
\end{enumerate}

\subsubsection{Results} \label{app:theory_results}
Under the above assumptions, the node placement schemes obtain the following bias rates:

\begin{itemize}
    \item \textbf{Randomly:} $\text{Bias} \approx \frac{L^2}{24}  \frac{1}{2\pi} \int_0^{2\pi} \nabla^2 f(p(\theta)) + \mathbb{E}_c[\epsilon(g_k)] \, d\theta  = O(L^2) + \Delta \epsilon$ 
    
    \item \textbf{Systematic:} $\text{Bias} \approx \frac{1}{2\pi} \int_0^{2\pi} \nabla f(p(\theta)) \cdot \delta+ \epsilon(g_k(\theta)) d\theta  = O(L) + \Delta \epsilon$ 
\end{itemize}

This shows that randomly placed nodes incur less bias $O(L^2)$ than systematically placed nodes $O(L)$. The intuition shown in the math above is that the expectation of the offset allows the first order term in the Taylor expansion to cancel out.

\paragraph{Increasing density }
For fixed $d$, as $L \to 0$ (density increases):
\begin{itemize}
    \item Randomly: $O(L^2) \to 0$ as $L \to 0$. Because $\Delta \epsilon$ is assumed to be $o(L^2)$, the noise bias also vanishes faster than the main term.
    \item Systematic: $O(L) \to 0$ as $L \to 0$. Because $\Delta \epsilon$ is assumed to be $o(L^2)$, the noise bias also vanishes faster than the main term.
\end{itemize}

Thus, bias for randomly placed nodes decays faster than systematically.

\subsection{Bias in systematic placement decays with increasing $d$} \label{app:theory_decay}

We previously showed that the bias from systematic node placement scaled with $\delta$, a worse rate than randomly placed nodes. However, in this section we show that when $d \to \infty$, even this term in the bias decays. Specifically, we show that the offsets $\delta_x$ and $\delta_y$ averaged across $\theta$ decay as the radius $d$ increases. 

\subsubsection{Setup}
We define the average offset $\overline{\delta(c, d)}$ as:
\begin{equation}
\overline{\delta(c, d)} = \frac{1}{2\pi} \int_{0}^{2\pi} \delta(c, d, \theta) \, d\theta
\end{equation}
where $\delta(c, d, \theta)$ is the offset for a particular angle $\theta$, center $c$, and radius $d$. Our goal is to show:
\begin{equation*}
\overline{\delta(c, d)} \to 0 \quad \text{as} \quad d \to \infty
\end{equation*}

\subsubsection{Finding $g_k$}
For a given point $p$ on the circle, we find the corresponding grid point $g_k$. The point $p$ is given by:
\begin{equation}
p = c + r(\theta, d) = \begin{pmatrix} c_x \\ c_y \end{pmatrix} + \begin{pmatrix} d \cos\theta \\ d \sin\theta \end{pmatrix}
\end{equation}
The grid point $g_k$ is the center of the cell containing $p$:
\begin{equation}
g_k = \left( L \left\lfloor \frac{p_x}{L} \right\rfloor + \frac{L}{2}, \quad L \left\lfloor \frac{p_y}{L} \right\rfloor + \frac{L}{2} \right)
\end{equation}
To see this, note that $\left\lfloor \frac{c_x + d\cos\theta}{L} \right\rfloor$ is the number of cells the point $p$ is from the origin in the $x$ direction. Multiplying by $L$ locates the bottom left location of the cell $p$ lies in. Adding $L/2$ locates the middle of the cell. 

We focus on the $x$-coordinate (the same logic applies to $y$). Substituting $p_x = c_x + d\cos\theta$:
\begin{equation}
g_{k,x} = L \left\lfloor \frac{c_x + d\cos\theta}{L} \right\rfloor + \frac{L}{2}
\end{equation}

\subsubsection{Offset $\delta$}
The offset is defined as $\delta_x = g_{k,x} - p_x$:
\begin{align}
\delta_x(c, \theta, d) &= \left( L \left\lfloor \frac{c_x + d\cos\theta}{L} \right\rfloor + \frac{L}{2} \right) - (c_x + d\cos\theta) \\
&= L \left( \left\lfloor \frac{c_x + d\cos\theta}{L} \right\rfloor - \frac{c_x + d\cos\theta}{L} \right) + \frac{L}{2}
\end{align}

Let $u = \frac{c_x + d\cos\theta}{L}$. Then:

\begin{align}
\delta_x &= L \left( \lfloor u \rfloor - u + \frac{1}{2} \right) \\
&= L \left( \frac{1}{2} - \{u\} \right)
\end{align}
where $\{u\} = u - \lfloor u \rfloor$ denotes the fractional part.

\subsubsection{Taking the average over $\theta$}
Let $h(u) = \frac{1}{2} - \{u\}$. Note that as $\theta$ changes, $h(u)$ goes through phases; it is periodic and averages to zero over one period. To see this, we can think about moving along the circle in just the positive $x$-direction as we have before. As we move from $0$ to $L$, the offset is positive to the left of $g_k$, then zero, then negative. It repeats as we move from from $L$ to $2L$, and so on. The period of the $h(u)$ function is, therefore, $L$ and the average across one period is $0$. 

Now we compute the average $\bar{\delta}$:
\begin{equation}
\overline{\delta(c, d)}= \frac{L}{2\pi} \int_{0}^{2\pi} h\left( \frac{c_x + d\cos\theta}{L} \right) \, d\theta = \frac{L}{2\pi} \int_{0}^{2\pi} h\left( d \frac{\cos\theta}{L} + \frac{c_x}{L} \right) \, d\theta
\end{equation}

Since $\frac{c_x}{L}$ is a constant, we have:
\begin{equation}
\overline{\delta(c, d)} = \frac{L}{2\pi} \int_{0}^{2\pi} h\left( d \frac{\cos\theta}{L} + \text{constant} \right) \, d\theta
\end{equation}

Following the generalization of the Riemann-Lebesgue Lemma from \citet{siretki1982mathematical}:
\begin{equation}
\lim_{d \to \infty} \overline{\delta(c, d)} = \frac{L}{2\pi} \int_{0}^{2\pi} \bar{h} \, d\theta
\end{equation}
where $\bar{h}$ is the average of $h(u)$ over one period. Since $\bar{h} = 0$:
\begin{equation}
\lim_{d \to \infty} \overline{\delta(c, d)} = 0
\end{equation}

We can show the same steps for the $y$-coordinate. 

\subsubsection{Results}

From the results of the Taylor expansion in Section \ref{app:theory_results}, we could have observed that for a fixed grid size $L$, bias decreases with increasing $d$ if the gradient $\nabla f$ decay with distance (e.g., $f(d) = 1/(1+d)$).

However, the result from the generalized Riemann-Lebesgue Lemma is even stronger. We have shown that for \textit{any} effect function $f$ with a bounded gradient, the systematic bias decays to zero as $d \to \infty$. This is because the bias is proportional to the average offset $\overline{\delta(c, d)}$ (via $\delta$ term dominating Taylor expansion), and as we have proven:
\begin{equation}
\lim_{d \to \infty} \overline{\delta(c, d)} = 0
\end{equation}
Consequently, even if the gradient $\nabla f$ remains constant and does not decay with distance, the bias (for a given center) will still vanish. Intuitively, as the radius $d$ increases, the circle path passes through an increasing number of grid cells and the local offsets $\delta$ cancel each other out. 

This result is primarily relevant for the systematic bias. As we previously showed, with random node placement, the first order $\delta$ term vanishes due to the randomization.

\subsection{An example illustrating a couple of points}
Below shows three circles centered at the origin of varying radii: $L/2$, $L/\sqrt{2}$, and $L$. There are four square cells of length $L$, and their centers are denoted by black dots.

\begin{figure}
    \centering
    \includegraphics[width=0.75\linewidth]{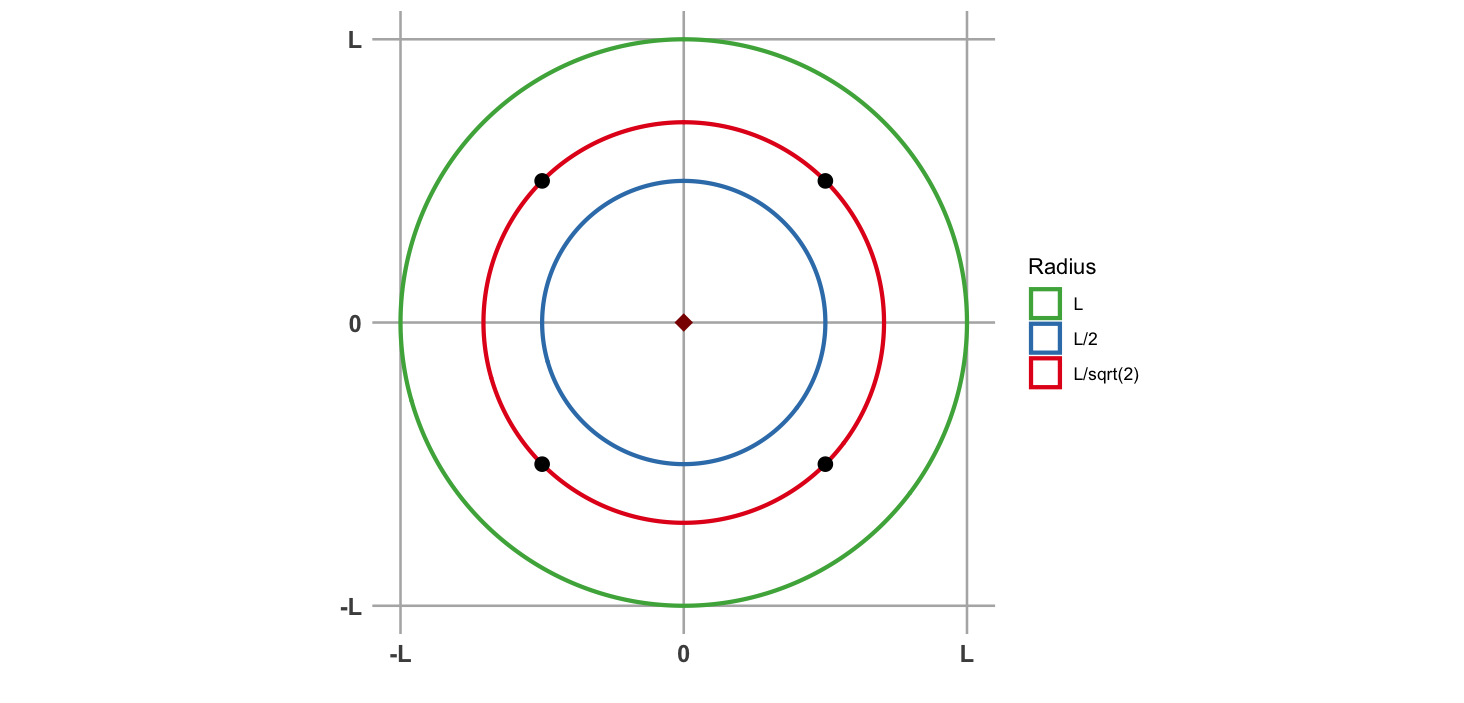}
\end{figure}

Assume the true effect $f$ decays rapidly with distance from the center. This is a case where the radius of the circle is large relative to the grid size and gradient of $f$ and results from the Taylor expansion do not hold. 

This picture illustrates a couple of points: 
\begin{enumerate}
    \item The circle average will be underestimated when $d = L/2$ and overestimated when $d = L$
    \item At $L/\sqrt{2}$, the circle average is (almost) perfectly estimated. This is not true generally but can happen in certain scenarios. If we do not assume any error terms $\epsilon_x$, the estimated circle average $\hat\mu = \mu$ exactly. 
    \item This illustrates that if all nodes are systematically placed at the corner of four cells, the error structure will be the same for all nodes. There will be no chance for errors to cancel across nodes (see section below on Error in causal estimators). 
    \item Circle averages are hard to estimate at small $d$ because there circle passes through less cells. In this case, only four cells are passed through. 
\end{enumerate}

\subsection{Error in causal estimators} \label{app:theory_causal}

Now we show how estimation error in the circle averages propagates to estimation error and bias in the AME estimators. For exposition, assume a Bernoulli random design where $Z_i \sim \text{Bern}(p)$, though results hold for any design. 

\subsubsection{Estimation error}

The oracle Horvitz-Thompson estimator for the AME assuming we observed the true circles averages, $\mu_i$, is denoted $\hat\tau^*$. 

$$\hat\tau^* = \frac{1}{Np} \sum_{i=1}^N Z_i \mu_i(Y,d) - \frac{1}{N(1-p)} \sum_{i=1}^N (1-Z_i) \mu_i(Y,d)$$

Note that as in \citet{Wang2025aoas}, circle averages are function of the observed data $Y$ and distance, $d$. 

In practice, we have to use the estimated circle averages, $\hat\mu_i$ to obtain the Horvitz-Thompson estimator AME, denoted $\hat\tau$. 

$$\hat \tau = \frac{1}{Np} \sum_{i=1}^N Z_i \hat\mu_i(Y,d) - \frac{1}{N(1-p)} \sum_{i=1}^N (1-Z_i) \hat\mu_i(Y,d)$$

Without loss of generality, the estimated circle averages can be represented as the true circle average and some error term as $\hat\mu_i(Y, d) = \mu_i(Y, d)  + \tilde\epsilon_i(Y, d) $. That is, $\tilde\epsilon_i$ denotes the error in the estimated circle averages (as opposed to the error in a particular point). 

Then the difference in the oracle estimator and the estimator with modeled outcomes is: 
$$\hat \tau^* - \hat\tau = \frac{1}{Np} \sum_{i=1}^N Z_i \tilde\epsilon_i(Y,d) - \frac{1}{N(1-p)} \sum_{i=1}^N (1-Z_i) \tilde\epsilon_i(Y,d)$$

As the above discussion illustrates, the above expression will generally be non-zero because the estimation error in circle averages will not cancel. Even without interference, this estimation error can be extreme. For example, consider circles that are systematically placed and are far apart. Treated nodes have $f(d) = \frac{1}{1 + d}$ while the control nodes have $f(d) = 0$. From the Taylor expansion, we saw that bias for a single placed node scales with $\nabla f(p(\theta))$. In this example, treated nodes have gradient of $1/d^2$ while the control node has gradient $0$, causing treated nodes to have more estimation error than control nodes. 

The estimation error of the AME also depends on $L$, $d$, and the function $f$. Using previous arguments, we expect the estimation error to decrease with increasing density and radius under certain conditions. 

\subsubsection{Bias under design-based inference}

Without loss of generality, assume that the potential circle average outcome for node $i$ with fixed centers can be written as:

$$\mu_i(\mathbf{Y}(\mathbf{Z}), d) = \hat\mu_i(\mathbf{Y}(\mathbf{Z}), d) + \tilde\epsilon_i(\mathbf{Y}(\mathbf{Z}), d)$$

where $\tilde\epsilon_i(\mathbf{Y}(\mathbf{z}), d)$ is the ``potential error," the modeling error in the circle average as a function of potential outcomes. The observed error is $\tilde\epsilon_i(Y,d) = \tilde\epsilon_i(\mathbf{Y}(\mathbf{z}), d) I(\mathbf{Z} = \mathbf{z})$

To get the bias under design-based inference, we take the expectation of the estimation error across $Z$:

$$\mathbb{E}_\mathbf{Z} \left\{\frac{1}{Np} \sum_{i=1}^N Z_i \tilde\epsilon_i(\mathbf{Y}(\mathbf{Z}), d) - \frac{1}{N(1-p)} \sum_{i=1}^N (1-Z_i) \tilde\epsilon_i(\mathbf{Y}(\mathbf{Z}), d)\right\}$$

In almost all cases, this bias will be non-zero because modeling induces different errors in the treated and control nodes. Even the Horvitz-Thompson estimator accrues bias from the outcome modeling. 

By linearity of expectations, the bias can be rewritten as: 

$$ \frac{1}{N} \sum_{i=1}^N \left\{ \mathbb{E}_\mathbf{Z}\left[\tilde\epsilon_i(\mathbf{Y}(\mathbf{z_{-i}}), d)|z_i = 1\right] -  \mathbb{E}_\mathbf{Z}\left[\tilde\epsilon_i(\mathbf{Y}(\mathbf{z_{-i}}), d)|z_i = 0\right] \right\}$$

This form clearly shows that the bias does not decay with increasing number of nodes, $N$. As the number of nodes increases, the spatial domain grows at a fixed density. Only density impacts the bias, not the number of nodes. 

The same ideas hold when we assess the bias across node placements as we take another expectation with respect to node centers, $c$. In this case, the bias decays with density as $O(L^2)$, as opposed to $O(L)$ with systematically placed nodes.

\section{Simulations}
\subsection{Additional Simulations} \label{otherdgp}
We include simulations with two other data generating processes. Figure \ref{fig:sim_random} contains results from a null effect data generating process. Under this specification, $AME(d) = 0$ for all $d$. The observed outcome at location $x$ is a standard Gaussian random variable: $Y_x(\mathbf{z}) \sim N(0,1)$. 

\begin{figure}
    \centering
    \includegraphics[width=1\linewidth]{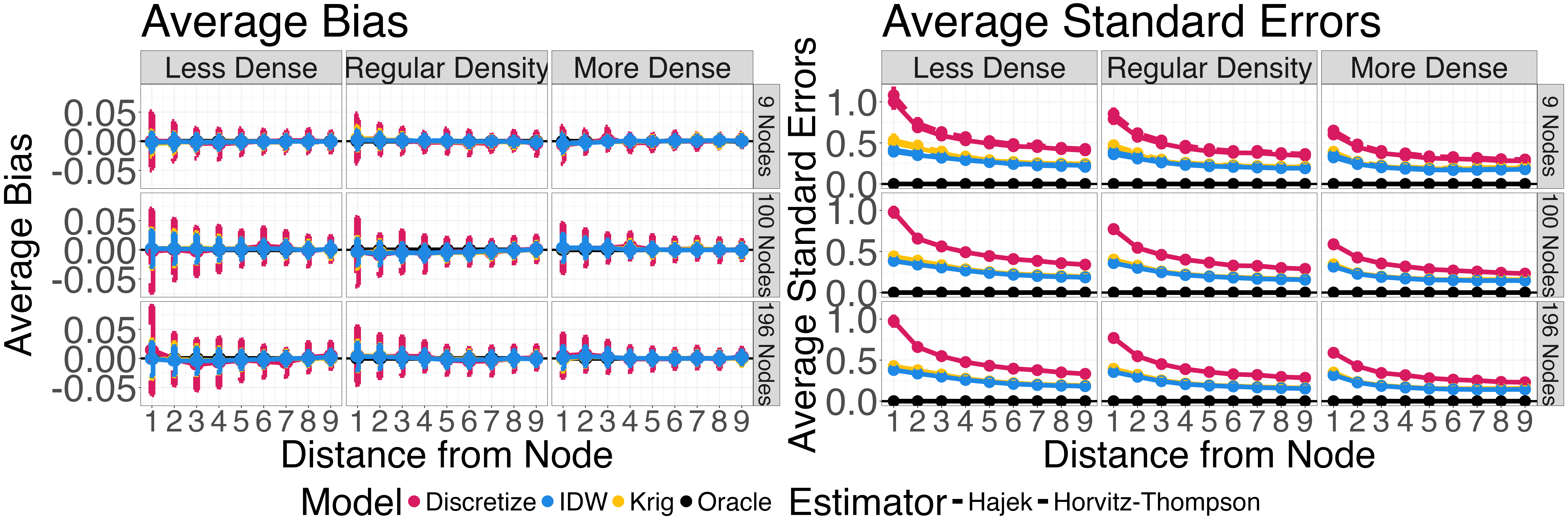}
    \caption{Simulation results with null effect data generating process. randomly placed nodes. Average bias is across 50 random node placements, and error bars capture variation across node placements. Facet columns vary outcome density, while facet rows vary number of nodes. Three outcome models and two estimators are presented. }
    \label{fig:sim_null}
\end{figure}

The null effect data generating process demonstrates the relationship between standard errors and distance. Since the effect is zero everywhere, standard errors are only driven by the amount of observed data used to estimate the circle averages. With increasing distance, standard errors decrease because more observed outcomes are being used for estimation as shown in Figure \ref{fig:vary-d}. 

\begin{figure}
    \centering
    \includegraphics[width=1\linewidth]{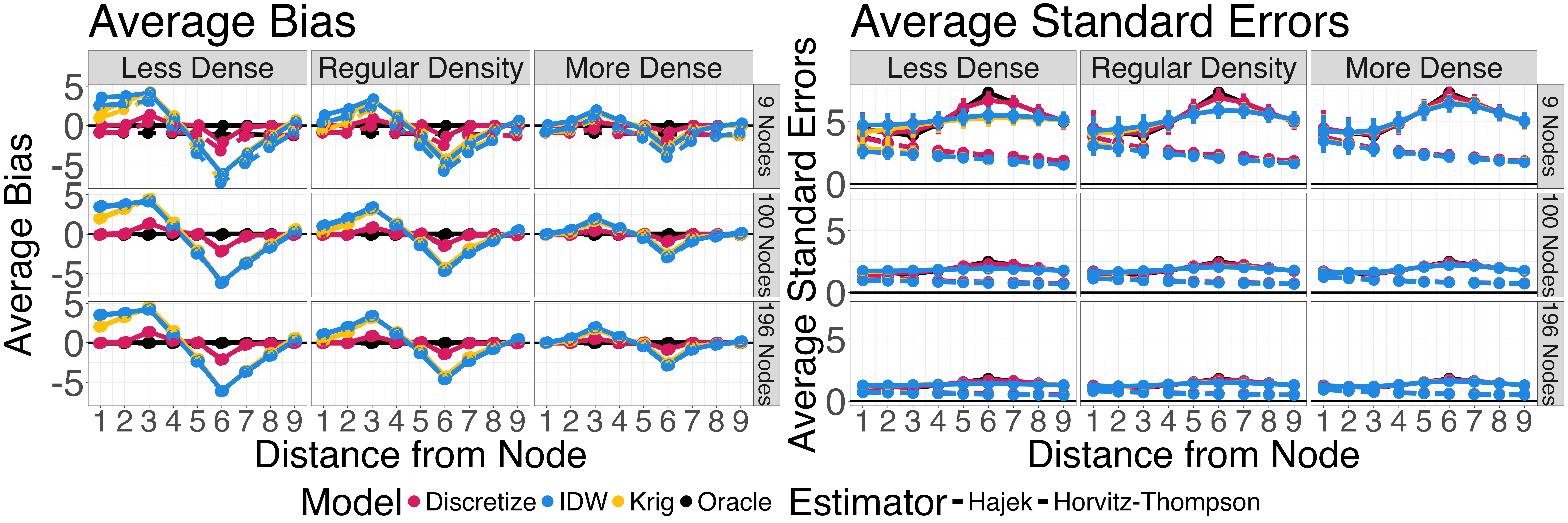}
    \caption{Simulation results with additive displacement data generating process. randomly placed nodes. Average bias is across 50 random node placements, and error bars capture variation across node placements. Facet columns vary outcome density, while facet rows vary number of nodes. Three outcome models and two estimators are presented. }
    \label{fig:sim_displacement}
\end{figure}

We additionally simulate a \textit{spatial additive displacement effect}. This process models displacement of outcomes to areas farther from intervention nodes. In this setting, potential outcomes are generated:

\begin{equation}
Y_x(\mathbf{Z}) = Y_x(0) + \sum_{i=1}^N \text{Pois} \left(\lambda(d_{ix})\right)Z_i  \mathbb{I}\{d_{ix} \leq 10\}
\end{equation}

\noindent where $ \lambda(d_{ix})$ is:

$$ \begin{cases}
0 & \text{if $d_{ix} \leq 3$} \\
4d_{ix}  - 12 & \text{if $d_{ix}  \in (3, 6]$} \\
-3d_{ix}  + 30 & \text{if $d_{ix}  \in (6,10)$} \\
0 & \text{otherwise}
\end{cases}$$ We note that the effect is 0 for distances beyond 10 units from intervention nodes to ensure that the local interference assumption posed in \citet{Wang2025aoas} holds (See Section \ref{sec:local-interference}).  Figure \ref{fig:sim_displacement} contains results under this data generating process, which mirror findings from the main text.  

\subsection{Impact of node placements} \label{placements}

\begin{figure}
    \centering
    \includegraphics[width=1\linewidth]{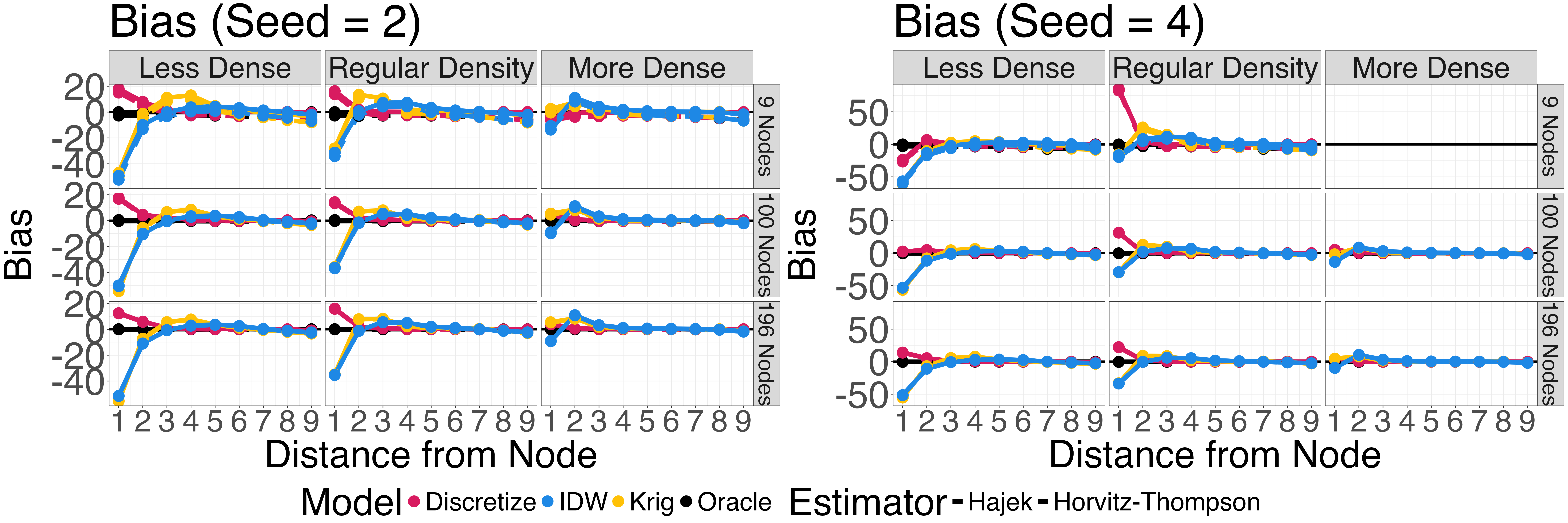}
    \caption{Simulation results for two random node placements determined by random seed. Facet columns vary outcome density, while facet rows vary number of nodes. Three outcome models and two estimators are presented.}
    \label{fig:sim_seeds}
\end{figure}

Figure \ref{fig:sim_seeds} shows that the randomness in the node placements can result in different performance of the outcome models. These differences can be especially large for settings with low number of nodes and under the discretized model.

\section{Application}
\subsection{Details on donut model} \label{donut}
We considered using a donut with equal weighting as an outcome model. We estimate the circle average at a distance $d$ by taking the average of all observed outcomes that fall within the donut, i.e. of all points within a distance of $d - h$ and $d + h$ of the intervention node, with $h$ chosen using cross-validation.

Since we only have outcomes at each street segment, we do not have enough observed outcomes for the donut approach to be viable at the distances we care about. Although the outcome space is relatively dense overall, at the distances we care about, they are sparse. To test this, we took a random sample of 100 hotspots and found the distance from all street segments to the hotspots. We only used 100 hotspots for ease of computation. We found that for $d = 100$ meters, we need $h$ to be at least 70 meters to get a median of 32 street segments within the 100 donuts. Having a donut be 140 meters wide to estimate effects at $d = 100$ is unreasonable and impossible. For this reason, we are unable to use the donut approach.

\subsection{Hyperparameter tuning} \label{disc-hyper}
The discretized model requires tuning of the resolution of the grid. We have found that there exists a tradeoff between accuracy and missingness in discretized models. For example, though very accurate, a fine grid of 300 x 300 suffers from missingness. Since the mid-point of the street segments are not located on a grid, many of the raster grids do not contain an observed outcome, and the model will be unable to produce an estimate for any point in those grids. Through exploration, we have found that a grid of 100 x 100 is reasonable for our setting. This grid does not suffer from much missingness and at the same time does not overly smooth.

For inverse distance weighting, we find through cross-validation that the power of the inverse should be one, and at minimum 2 points and at maximum 20 points located near the point of interpolation should be used.

\subsection{Direct effects $(d= 0)$}\label{app:d0}
Table \ref{tab:direct_effects} includes AME estimates at $d=0$ for each outcome type. These are equivalent to the direct effect of policing at hot spots. As in the main text, these results use inverse distance weighting as the outcome model and generate two sets of confidence intervals. 

\begin{table}[ht]
\centering
\begin{tabular}{lrll}
  \toprule
Outcome & Estimate & Conley 95\% CI & Permutation 95\% CI \\ 
  \midrule
Car and Motorcycle Thefts & -0.03 & (-0.103, 0.049) & ( -0.1, 0.047) \\ 
  Personal Robberies & -0.16 & (-0.431, 0.119) & (-0.552, 0.162) \\ 
  Homicides & 0.02 & (-0.009, 0.043) & ( -0.004, 0.038) \\ 
  Assaults & -0.00 & (-0.076, 0.073) & ( -0.067, 0.061) \\ 
   \bottomrule
\end{tabular}
\caption{Direct Effects by Outcome using IDW model as in main text. } 
\label{tab:direct_effects}
\end{table}

Like \citet{collazos2021hotspot} we find that the direct effects are insignificant, though the point estimates and standard errors not exactly identical to theirs. The primary reason is that \citet{collazos2021hotspot} use a fixed effects model and adjust for covariates while our analyses does not. However, the sign and significance of our direct estimates is consistent with the original paper.

\subsection{Robustness to $\bar d$ for Conley spatial HAC standard errors}\label{conley-robust}

\begin{figure}[ht]
    \centering
    \includegraphics[width=0.75\linewidth]{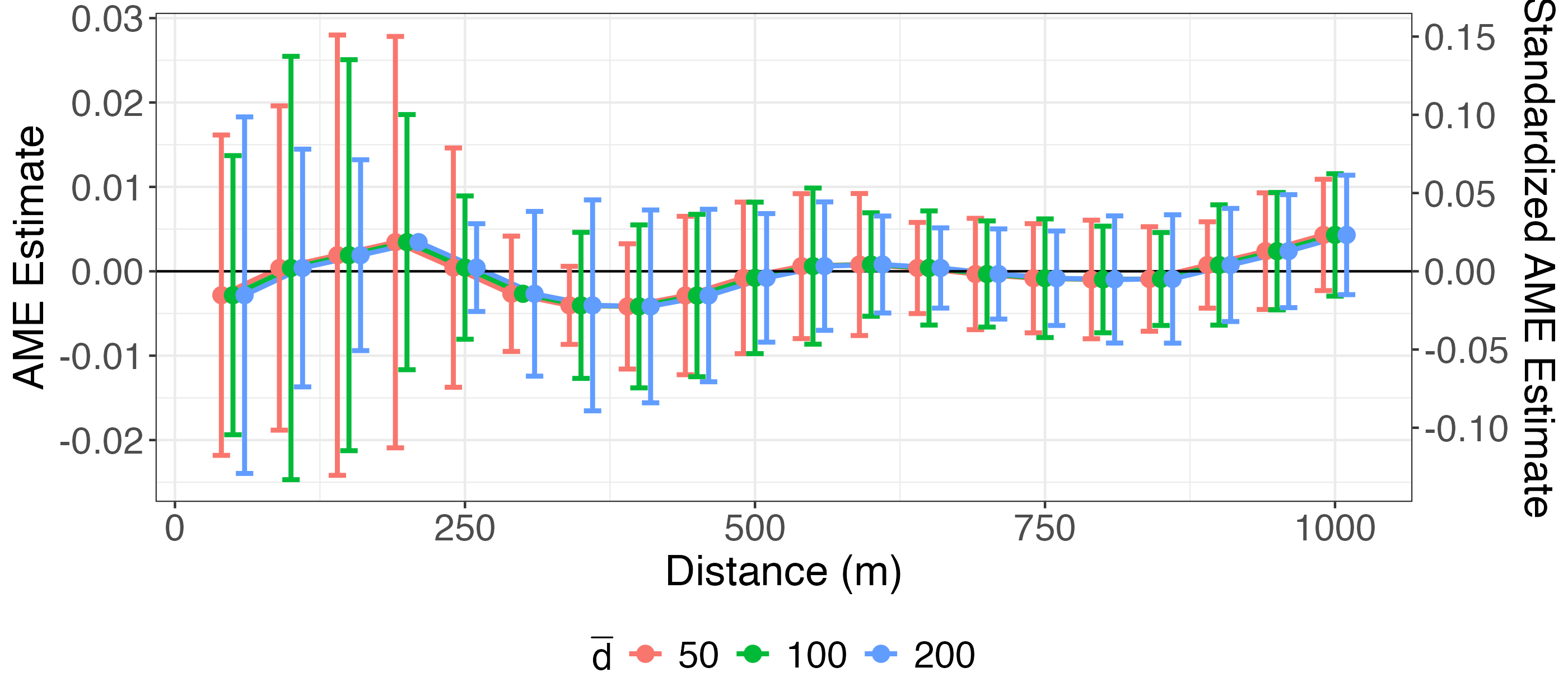}
    \caption{Varying the cutoff, $\bar d$, for estimation of Conley standard errors as suggested by \citet{Wang2025aoas}.}
    \label{fig:varying-dbar}
\end{figure}

Estimation of Conley spatial HAC standard errors as suggested by \citet{Wang2025aoas} requires making an assumption about the extent of interference. Specifically, \citet{Wang2025aoas} suggest assuming a cutoff value, $\bar d$: intervention nodes have no influence on outcome points located more than $\bar d$ away from them. Without a strong substantive basis to select this cutoff value, \citet{Wang2025aoas} suggest estimating the standard errors with varying $\bar d$ to assess the robustness of inference. In Figure \ref{fig:varying-dbar}, we present 95\% confidence intervals generated from the Conley standard errors using various choices for $\bar d$. Specifically, we look over $\bar d \in \{50, 100, 200\}$. \citet{collazos2021hotspot} assume that short-range spillovers are within 150m, so we believe these values for $\bar d$ are reasonable. 

We find that with increasing $\bar d$, the standard errors increase as expected. The results with $\bar d = 100$, which we present in the main analysis, are most similar to the intervals from inverting the permutation test. Our results are robust to these reasonable choices of $\bar d$; no matter the choice of $\bar d$, the results are still insignificant using 95\% confidence intervals.

\subsection{Robustness to other models}\label{robustness-model}
\begin{figure}[ht]
    \centering
    \includegraphics[width=0.75\linewidth]{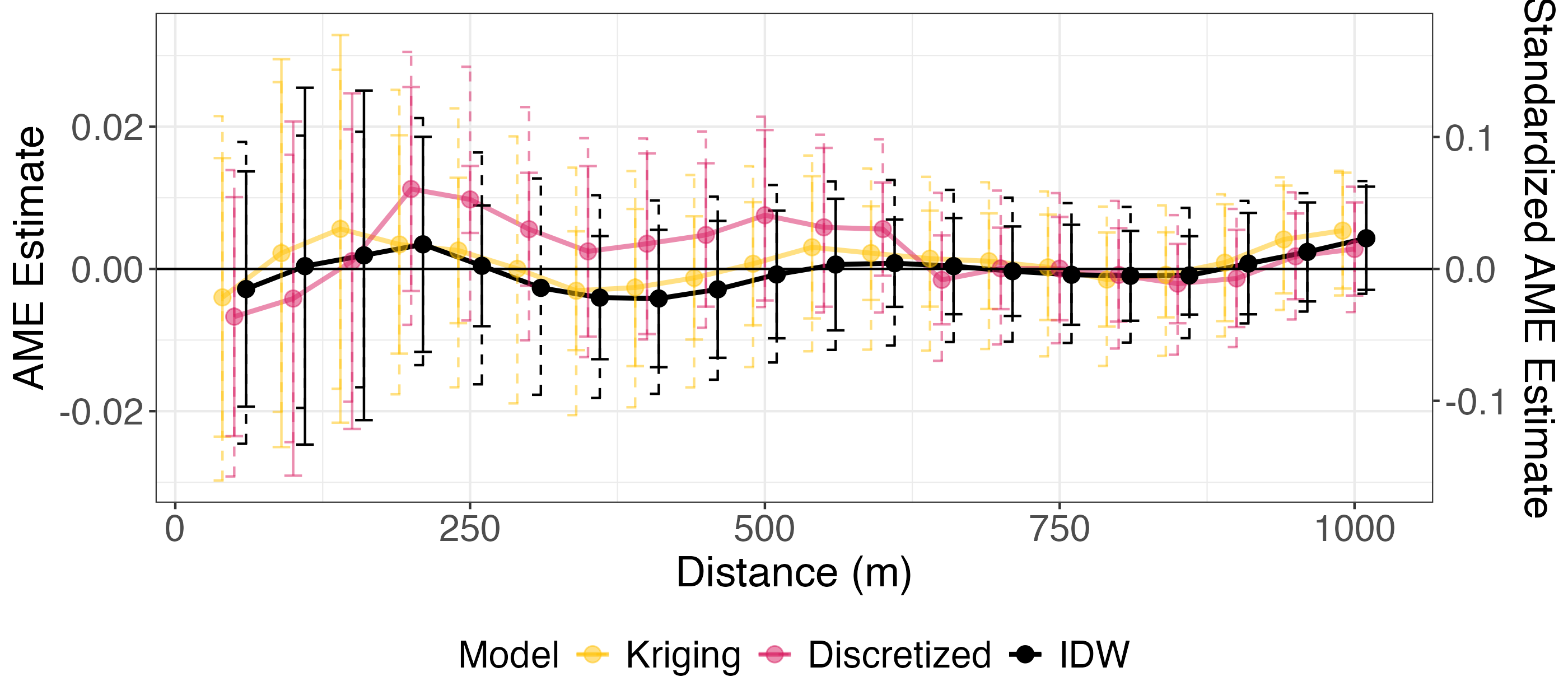}
    \caption{H\'ajek AME estimates using other modeling choices. 95\% confidence intervals are presented from estimation of the Conley Spatial HAC standard errors (solid) and inversions of permutation tests (dashed).}
    \label{fig:med-other-models}
\end{figure}

Although these results use IDW modeling for the outcomes, we note that our findings are robust to the choice of outcome model. Figure \ref{fig:med-other-models} shows the AME estimates under the other two models under consideration, kriging and the discretized model. Under these modeling choices, the AME estimates are still low in magnitude and insignificant at all distances. Our chosen IDW model has the most smooth AME curve, which may additionally suggest that this model best fits the data. 

\end{refsection}
\end{document}